\documentclass{article}

\usepackage{arxiv}

\usepackage[tbtags]{amsmath}
\usepackage{amssymb}
\usepackage{fancyhdr}
\usepackage{tabularx}
\usepackage{booktabs}
\usepackage{graphicx}
\usepackage{subfigure}
\usepackage{wrapfig}
\usepackage{multicol}
\usepackage{verbatim}
\usepackage{mathrsfs}
\usepackage{multirow}
\usepackage{siunitx}
\usepackage{xcolor}

\usepackage{amsfonts}%
\usepackage{amsthm}%
\usepackage{textcomp}%
\usepackage{manyfoot}%
\usepackage{algorithm}%
\usepackage{algorithmicx}%
\usepackage{algpseudocode}%
\usepackage{listings}%

\usepackage{dcolumn}

\usepackage{mathptmx}
\usepackage{etoolbox}
\usepackage{caption}
\usepackage{epstopdf,epsfig}
\usepackage{amsbsy}
\usepackage{enumitem}
\usepackage{subcaption}
\usepackage{xr-hyper}
\usepackage{hyperref}

\usepackage[utf8]{inputenc} 
\usepackage[T1]{fontenc}    
\usepackage{hyperref}       
\usepackage{url}            
\usepackage{booktabs}       
\usepackage{amsfonts}       
\usepackage{nicefrac}       
\usepackage{microtype}      

\usepackage{graphicx}
\usepackage{natbib}
\usepackage{doi}

\usepackage{bm}
\usepackage{caption}

\usepackage{epstopdf, epsfig}
\usepackage{amsthm}

\usepackage{amsmath}
\usepackage{accents} 
\usepackage{amssymb}
\usepackage{amsxtra}
\usepackage{enumitem}
\usepackage{bbold}

\usepackage{subcaption}
\usepackage{upgreek}
\usepackage{url}
\usepackage{cleveref}       

\newcommand{\dd}{\,\mathrm{d}}
\newcommand{\euler}{\mathrm{e}}

\newcommand{\stirlingii}{\genfrac{\{}{\}}{0pt}{}}

\crefname{section}{§}{§§}
\Crefname{section}{§}{§§}

\title{On a complete analytical solution of transient friction in pipe flow}

\newif\ifuniqueAffiliation
\uniqueAffiliationtrue

\ifuniqueAffiliation 
\author{ \href{https://orcid.org/0000-0002-8065-1449}{\hspace{1mm}F. Javier Garc\'ia Garc\'ia} \\
	Thermal Systems \& Heat Transfer  \\
	Research Group (SISTER)\\
	University of A Coru\~na\\
	Campus Industrial de Ferrol\\
	C/ Mendiz\'abal, s/n\\
	15403-Ferrol, A Coru\~na,
	Spain\\
	\texttt{f.javier.garcia.garcia@udc.es} \\
	\And
	\href{https://orcid.org/0000-0002-9598-5249}{\hspace{1mm}Pablo Fari\~nas Alvari\~no} \\
	Department of Naval \& Industrial Engineering\\
	Polytechnic School of Engineering \\
	University of A Coru\~na\\
	Campus Industrial de Ferrol\\
	C/ Mendiz\'abal, s/n\\
	15403-Ferrol, A Coru\~na,
	Spain\\
	\texttt{pablo.farinas@udc.es} \\
}
\else
\usepackage{authblk}

\author[1]{%
	\href{https://orcid.org/0000-0000-0000-0000}{\usebox{\orcid}\hspace{1mm}David S.~Hippocampus\thanks{\texttt{hippo@cs.cranberry-lemon.edu}}}%
}
\author[1,2]{%
	\href{https://orcid.org/0000-0000-0000-0000}{\usebox{\orcid}\hspace{1mm}Elias D.~Striatum\thanks{\texttt{stariate@ee.mount-sheikh.edu}}}%
}
\affil[1]{Department of Computer Science, Cranberry-Lemon University, Pittsburgh, PA 15213}
\affil[2]{Department of Electrical Engineering, Mount-Sheikh University, Santa Narimana, Levand}
\fi

\renewcommand{\headeright}{Scientific Article}
\renewcommand{\undertitle}{Scientific Article}
\renewcommand{\shorttitle}{Analytical solution of transient friction in pipe flow}

\hypersetup{
pdftitle={On a complete analytical solution of transient friction in pipe flow},
pdfsubject={physics.flu-dyn},
pdfauthor={F.J.~Garcia Garcia, P.~Fari\~nas Alvari\~no},
pdfkeywords={Turbulence, Turbulent flow, Transient friction},
}

\begin{document}
\maketitle

\begin{abstract}
The present research is a theoretical study about the transient friction created in circular pipe mean flow, whenever an incompressible Newtonian fluid is accelerated through a monotonously-increased mean-pressure gradient.
The resulting friction stress is the sum of two components, one laminar and the other purely turbulent, not synchronised between them.
Each component is analysed separately, in a series of theoretical experiments that explore various possibilities,
depending on the degree of asynchrony between them.
It is found that in some cases the transient friction is higher than in equal-$Re$ steady-sate flow,
but in some others it is noticeably lower.
This work provides an analytical explanation for most of the important and interesting phenomena reported in the literature.
To do so, it takes advantage of the Theory of Underlying Laminar Flow (TULF), already introduced in previous works of same authors.
The TULF predicts quite approximately what is observed in experiments, including the transient skin-friction coefficient and the presence of mean-velocity overshoots.
Additionally, the role of the time constant in turbulent mean flow is examined and related to the turbulence's frozen time.
Finally, a study of the logarithmic layer evolution in the transient flow is accomplished, which results destroyed during the increase of turbulence occurring along the transient.
In summary, the present work unveils new knowledge about transient friction in unsteady flows.
\end{abstract}


\section*{\small List of acronyms}\label{sct:acronyms}
  \vspace{-5mm}

\begin{multicols}{2}
\begin{footnotesize}
\begin{enumerate}
\item[CTD]Centreline turbulent dissipation
\item[DoF]Degree of freedom
\item[FTAM]Frozen-turbulence analytical model
\item[GAS]General analytical solution
\item[HSL]Hyperlaminar sublayer
\item[MBA]Mean bulk acceleration, 
$\dd \widetilde{u}/\dd \tau\mathop{=}\tfrac 12 \dd Re / \dd \tau$
\item[MPG]Mean-pressure gradient, $\Pi(\tau)$
\item[MWSS]Mean wall-shear stress, $\sigma_w(\tau)$
\item[PFT]Permanently-frozen turbulence
\item[PTC]Purely turbulent component (of mean flow), $u_T(\tau,\alpha)$
\item[RANSE]Reynolds-averaged Navier-Stokes equation
\item[RSS]Reynolds shear stress, $\sigma(\tau,\alpha)$
\item[RSSRG]Reynolds-shear-stress radial gradient, $\varSigma(\tau,\alpha)$
\item[S-]Steady-state (fully-developed mean steady-state)
\item[SDoF]Spatial DoF ($\Pi_i$, $\chi_i$, $q_i$, $\Delta \Pi\mathop{=}\Pi_2-\Pi_1$, $\widehat{\Pi}_i$, $\Delta\widehat{\Pi}\mathop{=}\widehat{\Pi}_2\mathop{-}\widehat{\Pi}_1$)
\item[TDoF]Temporal DoF ($\Delta \tau$, $\tau_0$, $\tau_2$, $\widehat{\Delta \tau}\mathop{=}\tau_2-\tau_0$)
\item[TULF]Theory of underlying laminar flow
\item[U-]Unsteady (fully-developed mean unsteady)
\item[ULF]Underlying laminar flow, $u_L(\tau,\alpha)$
\item[WMPG]Weighted mean-pressure gradient, $\widehat{\Pi}(\tau)$
\end{enumerate}
\end{footnotesize}
\end{multicols}

\section{Introduction}
\label{sct:intro}

The phenomenology of unsteady/transient mean flows (U-flows), particularly when they are turbulent, is overtly characterised by the appearance of non-conventional events,
which lead to the deformation of their mean-velocity profiles, when compared with those corresponding to equal-Reynolds-number steady-state mean flows (S-flows).
Said deformations, often quite striking, must occur in the temporal evolution of any U-flow, as demonstrated in \citep[Sect. 6.5]{GF22}.
The variety of deformed mean-velocity profiles is ample and one cannot avoid the impression that those reported in the comparatively few papers studying U-flows 
are just a small fraction of what could be uncovered if the research community initiated a systematic study of them.

This research is a continuation of \citep{GF24} and also explores the dynamics of transient friction, although this time applied to circular pipe flow,
which is known to undergo no secondary flows. 
In addition, its corresponding laminar flow is associated to pure shear, which could be understood as the canonical manner to address a formal study of friction. 
Consequently, this research will be concerned with the transient fully-developed mean flow in pipes, created when a steady-state mean flow (S-flow), characterized by Reynolds number ${Re}_1$, 
undergoes a monotonous mean-pressure gradient (MPG) increase, until it attains a new steady state with ${Re}_2 > Re_1$
(although very interesting, the case $Re_2<Re_1$ will not be considered). 
Examples of these flows can be found in previous experimental and DNS studies already referenced in \citep{GF24}.
Also, we shall take full advantage of the capabilities of our analytical approach since, in addition, 
this study will report a whole set of possible instances for the friction coefficient, depending on how fast the change of MPG is and how delayed is the associated production of new turbulence.

\begin{figure}
	\begin{center}
		\leavevmode
		\includegraphics[width=0.5\textwidth, trim = 0mm 0mm 0mm 0mm, clip=true]{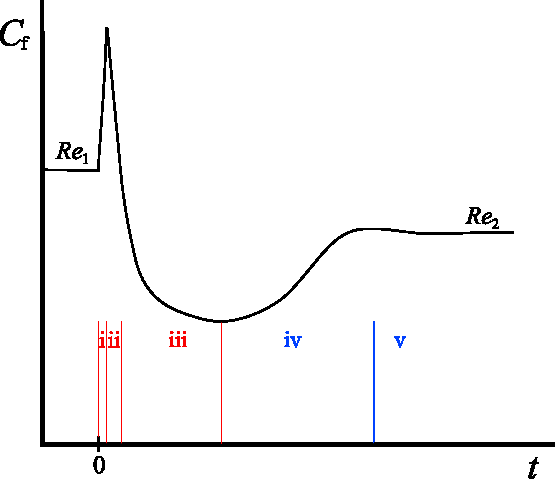}
		\caption{\small Idealised schematic view of the  `\textit{canonical}' bathtub.}
		\label{fig:bathtub}
	\end{center}
\end{figure}

This work will depart from the skin-friction coefficient $C_f(t)$ arising in a particular class of U-flows, which is ideally summarised in Fig. \ref{fig:bathtub} and already profusely described in \citep{GF24}.
It was shown in \citep{GF24} that this curve, herein called the `\textit{bathtub}', is the outcome of a rather complicated evolution and dynamics,
in which two transient volumetric forces act upon the fluid in a non-synchronised way, one pushing and the other pulling, with relative intensities that change with time in every spatial region. 
The bathtub is split into five stages, numbered i to v (limited by times $t_i$ to $t_v$), whose origin and causes were explained in \citep{GF24}.
Herein, an analogous explanation will be furnished for the five stages, but for pipe flow.
We shall see that two dynamical agents are needed, acting non-synchronously, to explain each of the bathtub features.
This research will focus primarily on forces, because they alone can explain the observed dynamics.

The references given in \citep{GF24} will not be repeated here, although they remain valid for the present research.
A comprehensive study of transient friction for complex surfaces, \citep{PS09}, should also be added to this reference list, as it provides a rigorous and original perspective on the problem,
although it differs from our approach on the type of averages considered.
Furthermore, we cannot omit two particular works that study rapidly accelerated pipe flows, which report skin-friction curves akin to the bathtub, 
and contribute important data to understand transient friction phenomena.
The first reference, \citep{HSH16}, is a DNS of an extremely accelerated pipe flow, 
whose evolution from initial to final S-flows is almost identical to that described in \citep[Fig. 1]{HS13}: a quasi-Heaviside step.
The second, \citep{GLC21}, is an analogous study, though with a moderately rapid increase of bulk velocity, some of whose results will also be reproduced herein (in particular, the experiment TP4).
Regrettably, none reports the MPG generated during the transient flows, which is crucial data for any attempt to reproduce their results from an analytical viewpoint.
Section \ref{sct:explBluntPeak} will be devoted to analyse the consequences of such data omission.

The study of $C_f$ developed in the present research not only furnishes information about the friction drag force acting on the mean flow,
but also about the time constant associated to the evolution of U-flows.
The relationship between the flow's time constant and the evolution of turbulence is most interesting and revealing,
for the shape actually taken by $C_f(t)$ depends greatly upon the synchronisation of the forces causing the flow.
An approach with explicit mathematics, such as the one reported herein, provides an additional insight and the possibility of reaching general conclusions.

The methodology for this research is based on the so-called Theory of Underlying Laminar Flow (TULF, see \citep{GF20}).
The TULF explains mathematically any fully-developed mean flow occurring in a pipe as the outcome of two competing forces, one pushing and the other pulling.
The pushing force is due to active agents such as MPG and gravity, which cause the component of mean motion called the Underlying Laminar Flow (ULF).
The pulling force is due to a single reactive agent, the Reynolds shear-stress radial gradient (RSSRG), and causes the Purely Turbulent Component (PTC).
ULF and PTC are both much greater than the mean-velocity field we actually observe in a flow, which is the difference between ULF and PTC at every point.
ULF and PTC are not synchronised and sometimes one is significantly greater than the other, at least locally.
Those local differences of ULF and PTC explain, with great accuracy, the extraordinary deformations observed in the mean-velocity profiles of most U-flows.
This research follows the mathematical definitions of laminar and turbulent flows offered in \citep{GF25}.

Henceforth, the terms {U-profile}, U-field, U-flow, U-configuration... will be used to denote the transient/unsteady mean profile of any quantity,
or any characteristic of transient/unsteady mean flows, whereas the terms {S-profile}, S-field, S-flow, S-configuration... will refer to the corresponding steady-state mean profile or characteristic.
The word '\textit{mean}' always refers to '\textit{ensemble-average}'.
Fully-developed mean flows are always assumed. 
 
This is the eighth research article in a series that began with \citep{GF19a,GF19c,GF20,GF21,GF22,GF24,GF25}.
Certain familiarity with those works is recommended, in order to fully understand the scope and methods of this one.

\section{Nomenclature, conventions and background}\label{sct:background}

Most of the background needed for the present research has been already explained \citep{GF20,GF22}.
The studied mean flow is that developed in a circular pipe of radius $R$ and indefinite length.
A cylindrical coordinate system $(r,\theta,z)$ is assumed, whose $Z$ axis is coincident with the pipe's centreline.
The nomenclature, coordinate system, and dimensionless variables considered herein are identical to those of \citep{GF20}, with
\begin{equation}
 \label{eq:dimPP}
 \beta\mathop{=}\frac{z}{R} \ , \quad \alpha \mathop{=}\frac{r}{R} \in [0,1] \ , \quad \tau\mathop{=}\frac{t \nu }{R^2} \ , \quad
 u(\tau,\alpha)\mathop{=}\frac{U R}{\nu} \ , \quad p(\tau,\beta)\mathop{=}\frac{P R^2}{\rho \nu^2} 
\end{equation}
which hereinafter will be called the natural normalisation (or dimensionless natural units, defined formally by the convention $R\mathop{=}\rho\mathop{=}\nu\mathop{=}1$).
The dimensionless mean pressure gradient (MPG) and Reynolds shear stress (RSS) are given, respectively, by
\begin{equation}
 \Pi(\tau)\mathop{=} -\frac{\partial p}{\partial \beta}\mathop{=} -\frac{R^3}{\rho \nu^2}\frac{\dd P}{\dd z} \ , \qquad 
 \sigma(\tau,\alpha) \mathop{=} \frac{\langle u'_r u'_z \rangle R^2 }{\nu^2}
\end{equation}
We have recently discovered that the above normalisation was already proposed in \citep[Sect. 2.1]{DG04}.

In all equations, we consider ensemble-averaged quantities (called mean quantities) and we are exclusively concerned with mean flows,
not with individual realisations of any given flow (see \citep[Sect. 3.2.1.1]{Gar17}).
A Reynolds number ${Re}\mathop{=}2R\widetilde{U}/\nu$, with $\widetilde{U}$ the dimensional bulk velocity, is used everywhere.
Since hydraulic diameter equals geometric diameter for circular pipe, $D_h\mathop{=}2R$, obtained results can be readily compared with those of other geometries.
The Reynolds-averaged Navier-Stokes equation (RANSE) obtained from the natural normalisation adopts the form (see \citep[Eq. 2]{GF20})
\begin{equation}
\label{eq:RANSEPP}
 \frac{\partial u}{\partial \tau} - \left( \frac{\partial^2 u}{\partial \alpha^2}
 + \frac{1}{\alpha} \frac{\partial u}{\partial \alpha}\right)\mathop{=} \ \Pi - \frac{1}{\alpha}\frac{\partial (\alpha\sigma)}{\partial \alpha} \equiv \Pi(\tau) + \varSigma(\tau,\alpha)
\end{equation}
subject to the initial and boundary conditions 
\begin{equation}
\label{eq:boundCond}
	u(0,\alpha)\mathop{=}u_0(\alpha)  \ , \quad u(\tau,1)=0 \ , \quad \frac{\partial u(\tau,0)}{\partial \alpha}=0
\end{equation}
plus the dimensionless Reynolds-averaged continuity equation 
\begin{equation}
\label{eq:cont}
	\frac{\partial u}{ \partial \beta} =0
\end{equation}
which is trivial in this case (fully-developed mean flow).
In Eq. \eqref{eq:RANSEPP} the mean-velocity field $ u(\tau,\alpha)$ is at the left-hand side and the sources of mean motion (forces) at the right-hand side;
$\Pi(\tau)$ causing the underlying laminar flow (ULF) and $\varSigma(\tau,\alpha)$ causing the purely turbulent component (PTC).
The general solution of Eq. \eqref{eq:RANSEPP} is developed in the Hilbert space $L^2_{\alpha}(0,1)$, and was first reported in \citep{GF19a} (see Eq. \eqref{eq:equationGarcia} below).
Any mean flow driven by a MPG, $\Pi(\tau)$, and Reynolds shear-stress radial gradient (RSSRG), $\varSigma(\tau,\alpha)$), 
is bound to produce a mean-velocity field $u(\tau,\alpha)$ which,
thereby, would be expandable in a Fourier-Bessel series with the basis $\{ \phi_n(\alpha)\mathop{=}\sqrt{2}J_0(\lambda_n \alpha) /  J_1(\lambda_n)  \}$,
being $\lambda_n$ the zeroes of the Bessel function $J_0(x)$, \citep{GF19a}.

The universal time constant for fully-developed laminar circular-pipe U-flow, whose value is:
\begin{equation}
\label{eq:univCTE}
 \mathring{\tau}_c\mathop{=} \frac{t_c \, \nu}{R^2} \mathop{=}  0.165381775
\end{equation}
was first introduced in \citep[Sect. 7.2]{GF22}.  
For an energy-increasing system, such a time constant $\mathring{\tau}_c$ is defined as the dimensionless time needed for the dynamical system to reach the bulk velocity 
$\widetilde{u}_L(\mathring{\tau}_c) \equiv (1-\euler^{-1})\, \widetilde{u}_L(\infty) \approx 0.6321\, \widetilde{u}_L(\infty)$,
where $\widetilde{u}_L(\infty)$ is the limit steady-state response obtained after applying a unit Heaviside step at $\tau\mathop{=}0$ in the system's source 
($t_c$ in Eq. \eqref{eq:univCTE} is the actual time taken by the flow, as measured with a chronometer).
The response of a fully-developed laminar flow to any step increase in MPG is the same for all initial bulk velocities of such a flow or, otherwise put, 
for given pipe and fluid the time constant $\mathring{\tau}_c$ of a laminar flow is the same for all ${Re}$.
A similar interpretation is assumed when dealing with turbulent mean flows, except that in such cases the time constant is not universal, for it depends on the turbulence. 

This section is finished with a result that complements those reported in \citep[Sect. 3.6]{GLC21}.
In \citep[Eq. 3.13]{GF22} we proved that the exact mathematical relationship between the RSS and $C_f$ for S-flow is
\begin{equation}
\label{eq:CfRey}
 \int \limits_0^1 \dd \alpha \ \alpha \int \limits_{\alpha}^1 \dd \alpha' \ \sigma(\alpha') \mathop{=} \frac{{Re}}{4} \left( \frac{C_f {Re}}{16}-1 \right)
\end{equation}
After some algebra, Eq. \eqref{eq:CfRey} can also be expressed as
\begin{equation}
\label{eq:CfGF}
 C_f \mathop{=} \frac{16}{{Re}} + \left( \frac{8}{{Re}} \right)^2  \int \limits_0^1 \dd \alpha \ \alpha \int \limits_{\alpha}^1 \dd \alpha' \ \sigma(\alpha') 
\end{equation}
which is one of the most misleading equations of Fluid Mechanics, because one may be led to think that $C_f$ is the sum of a laminar contribution 
($C_{f_L}\mathop{=}16/{Re}$) and a purely turbulent contribution (the double integral), but it is not.
And it is not because $C_{f_L} \neq 16/{Re}$ when ${Re}$ does not correspond to laminar Hagen-Poiseuille flow, and ${Re}$ characterises a turbulent S-flow in Eq. \eqref{eq:CfGF}.
The true laminar contribution would be $C_{f_L}\mathop{=}16/{Re}_L\mathop{=}64/\Pi$, which corresponds to the ULF (check the Zero Theorem in \citep[Eqs. 2.8 and 3.6]{GF22}).
Eq. \ref{eq:CfGF} should be compared with \citep[Eq. 14]{FIK02} for fully-developed S-flow, which in their nomenclature reads
\begin{equation}
 \label{eq:FIK02}
 C_f\mathop{=}\frac{16}{{Re}_b}+32\int \limits_0^1 \overline{u'_r u'_z} \ r^2 \dd r
\end{equation}
However, in Eqs. \eqref{eq:CfRey} \& \eqref{eq:CfGF} $\sigma(\alpha)$ is the ensemble-averaged RSS over a denumerable set of realisations, 
whereas in Eq. \eqref{eq:FIK02} $\overline{u'_r u'_z}$ is an average around the azimuth coordinate.
They may not be compared directly.
On the other hand, both equations share the same deceit: the first term is not the laminar contribution.

\section{The frozen-turbulence analytical model (FTAM)}
\label{sct:model}

The frozen-turbulence analytical model (FTAM) will be the departing point for the development of this research. 
It contains the proviso that the RSS remains unchanged during a time interval after a monotonously-increased MPG has been applied to the U-flow,
that is, initially the mean flow evolves without changing the Reynolds stresses, which, after an interval, begins to increase at its own rate, different from that of the MPG.
The U-flow considered in this work is a horizontal circular pipe fully-developed mean flow generated upon applying a MPG $\Pi(\tau)$ and a RSSRG $\varSigma(\tau,\alpha)\mathop{=}-\alpha^{-1} \partial (\alpha \sigma) /\partial \alpha$,
the two sources (forces) of mean motion.
Most frequently, it will be a turbulent U-flow, meaning that it will have a nonzero PTC, $u_T(\tau,\alpha) \neq 0$.
The foundation of the TULF formalism is the general analytical solution (GAS) \eqref{eq:equationGarcia}, borrowed from \citep[Eq. 13]{GF20}
\begin{align}
\label{eq:equationGarcia}
u(\tau,\alpha)\mathop{=}&\sum \limits_{n\mathop{=}1}^{\infty} u_n(\tau) \ \phi_n(\alpha)\mathop{=}u_L(\tau,\alpha) + u_T(\tau,\alpha)\mathop{=}\sum \limits_{n\mathop{=}1}^{\infty} \left(u_{L_n}+u_{T_n} \right) \ \phi_n(\alpha)\mathop{=}  \sum \limits_{n\mathop{=}1}^{\infty} \left( u_{I_{L_n}}+u_{P_{n}}+u_{I_{T_n}}+u_{R_{n}} \right ) \phi_n(\alpha)\mathop{\mathop{=}} \nonumber \\
&\sum \limits_{n\mathop{=}1}^{\infty} \left[  u_{L_n}^{(0)}  \euler^{-\lambda_n^2 \tau}\ \mathop{+}  \frac{\sqrt{2}}{\lambda_n}
\int \limits_0^{\tau} \Pi(\tau')\ \euler^{-\lambda_n^2 (\tau-\tau')} \dd\tau' + u_{T_n}^{(0)} \euler^{-\lambda_n^2 \tau}\   \mathop{+} 
 \int \limits_0^{\tau} \varSigma_n(\tau')\ \euler^{-\lambda_n^2(\tau- \tau')} \dd\tau' \right] \phi_n(\alpha) \mathop{=} \notag \\
 & u_I(\tau,\alpha) + u_P(\tau,\alpha)+u_R(\tau,\alpha) 
\end{align}
with the names of IniTrans $u_I(\tau,\alpha)=u_{I_L}+u_{I_T}$, PresGrad $u_P(\tau,\alpha)$, RStress $u_R(\tau,\alpha)$, ULF $u_L(\tau,\alpha)$ and PTC $u_T(\tau,\alpha)$.
The IniTrans has its own ULF $u_{I_L}$ and PTC $u_{I_T}$.
The mean-velocity field $u(\tau,\alpha)=u_L + u_T$ also has ULF $u_L(\tau,\alpha)=u_{I_L}+u_P$ and PTC $u_T(\tau,\alpha)=u_{I_T}+u_R$.
In this paper, a '\textit{model}' is understood as the set of particular mathematical functions $u_0(\alpha)$, $\Pi(\tau)$ and $\varSigma(\tau,\alpha)$ 
inserted into Eq. \eqref{eq:equationGarcia}.

The dynamical system described by the FTAM is a Newtonian fluid inside a smooth circular pipe of diameter $2R$,
whose initial state is a S-flow of ${Re}_1$, which at $\tau\mathop{=}0$ is actuated upon by time-variable MPG $\Pi(\tau)$ and RSSRG $\varSigma(\tau,\alpha)$,
until the ensuing transient U-flow becomes stationary as a S-flow of ${Re}_2$.
Only fully-developed mean flows, i.e., ensemble-averaged flows, are considered.
As remarked in the Introduction, we shall limit to the case ${Re}_1<{Re}_2$ in this research.
The FTAM is fully characterised analytically in Apps. \ref{sct:modelS} \& \ref{sct:meanVelFieldS}, 
where the reader can find the explicit form of the functions $\Pi(\tau')$ and $\varSigma_n(\tau')$ appearing in Eq. \eqref{eq:equationGarcia}.
All results reported in this paper are obtained through direct calculation of those equations.

\section{The skin-friction coefficient for laminar U-flows}
 \label{sct:laminarFriction}  

The TULF permits the fractional study of U-flows by considering only isolated force fields causing them.
Thus, it is possible to analyse the evolution of laminar mean flows simply ignoring the RSSRG, i.e., switching off the turbulence.
Such an exercise will allow us to find out which properties of the bathtub depend exclusively on the laminar component of the mean flow.
For example, we shall see that the bathtub's initial sharp peak has no relationship with the turbulence.
Thereby, we can gain knowledge after studying the curve $C_f(\tau)$, or $C_f({Re})$, obtained under laminar conditions.

To fix ideas, the following definition of skin-friction coefficient will always be used, in the natural normalisation, for both laminar and turbulent U-flows:
\begin{equation}
 \label{eq:Cf}
 C_f(\tau) \mathop{=} \frac{2 \lvert \sigma_w(\tau) \vert}{\left[ \widetilde{u}(\tau) \right]^2} = \frac{8 |\sigma_w(\tau)|}{[{Re}(\tau)]^2}
\end{equation}
with $\widetilde{u}(\tau)\mathop{=}{Re}(\tau)/2$ and $\sigma_w(\tau)$ calculated from Eqs. \eqref{eq:bulkSeries} \& \eqref{eq:WSSsymmet}, respectively.
The laminar U-flow under study is the following: An initial laminar S-flow (Hagen-Poiseuille) of Reynolds number ${Re}_1$ is at $\tau\mathop{=}0$
subject to a monotonous MPG increase, given by Eq. \eqref{eq:AK+13PG}, during an interval $\Delta \tau$.
After this interval, the MPG remains constant and, some time later, the U-flow ends up converted into a laminar S-flow with ${Re}_2$.
The TULF can even simulate the case ${Re}_2\gg Re_1$, since no turbulence contribution is expected because no PTC will be included in the calculations.
The TULF allows such separation of laminar and turbulent components.

The theoretical experiments studied herein are listed in Table \ref{tab:laminarTE} and form two groups.
The driving force of U-flows, the MPG, is identical to that considered in \citep[Table 2]{GF24}, although the resulting $Re(\tau)$ are noticeably different,
since the flow geometry plays an important role in the transient evolution \textit{via} its time constant (another contribution of this research).
With such high ${Re}$, the actual flows would be turbulent, although only the ULF will be considered in each case.
It is advanced that the evolution of $C_f(\tau)$ is only interesting if $\Delta \tau \ll \mathring{\tau}_c \approx 0.165$, 
the universal time constant of laminar pipe flow.
Otherwise, the evolution is rather foreseeable and poses only moderate interest.
For water ($\nu\mathop{=}\SI{1.0533e-6} {\metre \squared \per \second}$ at $\SI{18}{ \degreeCelsius}$) and pipe radius ${R}\mathop{=}\SI{25}{mm}$, 
a dimensional time interval of $\Delta t\mathop{=} \SI{1}{\second}$ equals $\Delta \tau \mathop{=} 1.6853 \times 10^{-3}$ in the natural normalisation. 
Since $\SI{1}{\second}$ is already a rather short time for a typical valve to open (or a pump to attain a steady regime of pressure), 
the cases $\Delta \tau\mathop{=}10^{-3}$, $\Delta \tau\mathop{=}10^{-2}$ and $\Delta \tau\mathop{=}10^{-1}$ will be explored.
Calling $\Delta \tau$ the mean valve-aperture time is suitable, since such would normally be the operation executed in laboratory to increase the MPG of a flow.
The ULF for cases TE1-TE6 of Table \ref{tab:laminarTE} is calculated from Eqs. \eqref{eq:iniTransULF} \& \eqref{eq:PresGradBefore}-\eqref{eq:PresGradAfter},
and the bulk velocity and mean wall-shear stress (MWSS) from Eqs. \eqref{eq:bulkSeries} \& \eqref{eq:WSSsymmet}, without turbulent contribution.

\begin{figure}
\centering
	\begin{subfigure}[\footnotesize{$Re(\tau)$}]{\label{fig:ReLam1}
	    \includegraphics[width=70mm]{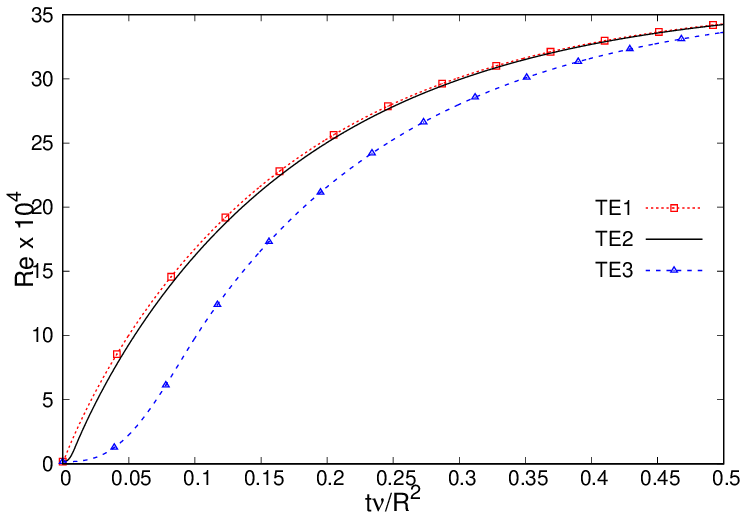}}
	\end{subfigure}
	\begin{subfigure}[\footnotesize{$Re(\tau)$}]{\label{fig:ReLam2} 
	    \includegraphics[width=70mm]{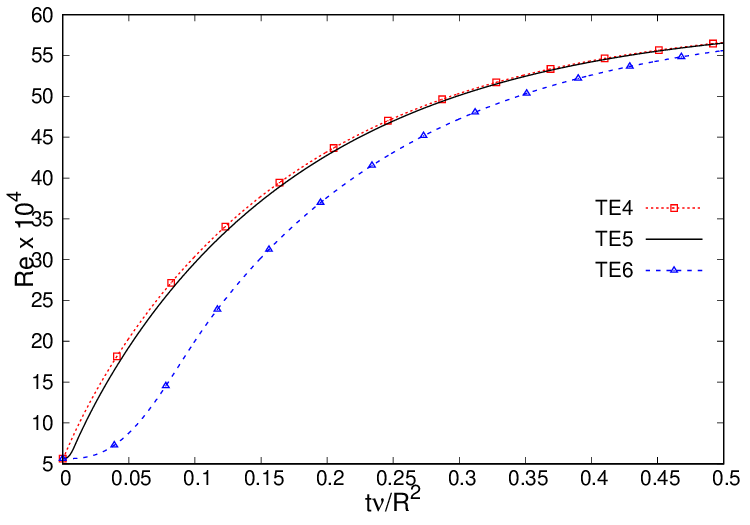}}
	\end{subfigure}
	\begin{subfigure}[\footnotesize{$|\sigma_w(\tau)|$}]{\label{fig:WSSLam1} 
	    \includegraphics[width=70mm]{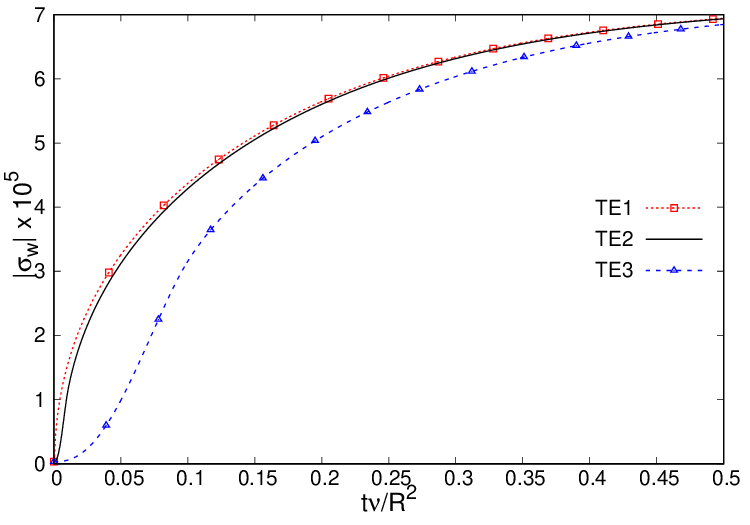}}
	\end{subfigure}
	\begin{subfigure}[\footnotesize{$|\sigma_w(\tau)|$}]{\label{fig:WSSLam2}
	    \includegraphics[width=70mm]{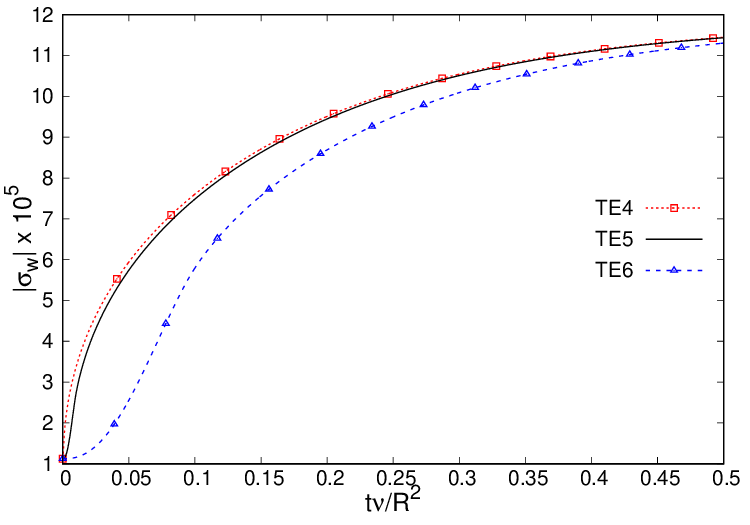}}
	\end{subfigure}
	\begin{subfigure}[\footnotesize{$C_f(Re)$}]{\label{fig:CfLam1} 
	    \includegraphics[width=70mm]{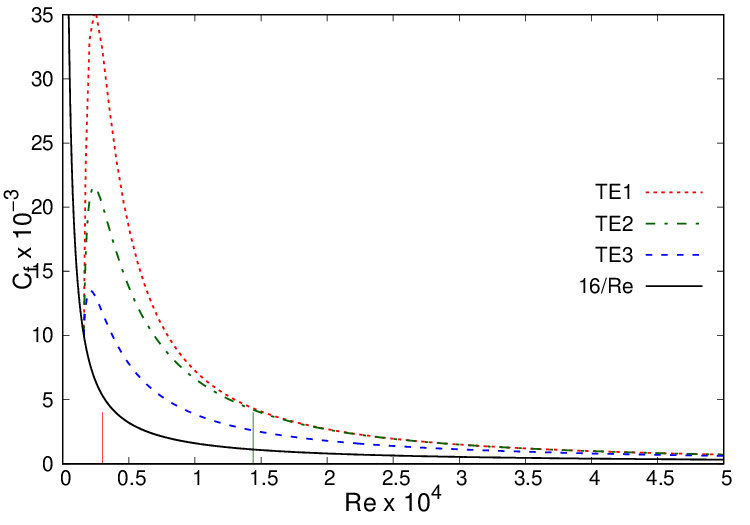}}
	\end{subfigure}
	\begin{subfigure}[\footnotesize{$C_f(Re)$}]{\label{fig:CfLam2}
	    \includegraphics[width=70mm]{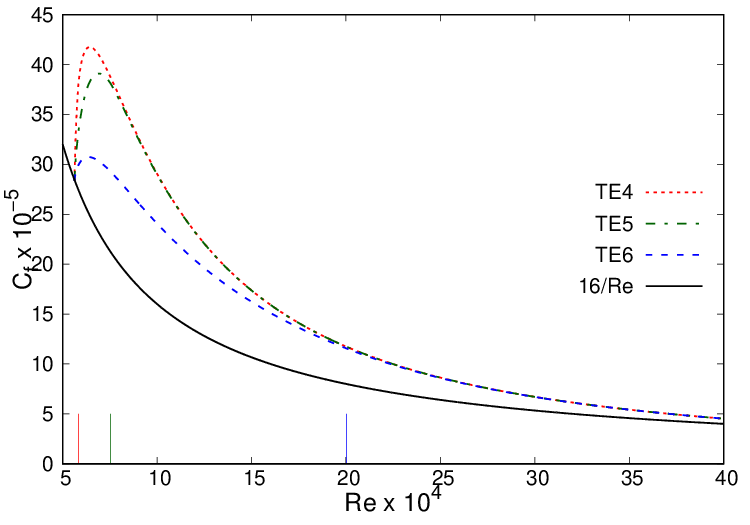}}
	\end{subfigure}
	\begin{subfigure}[\footnotesize{$C_f(\tau)$}]{\label{fig:CfLamT1}
	    \includegraphics[width=70mm]{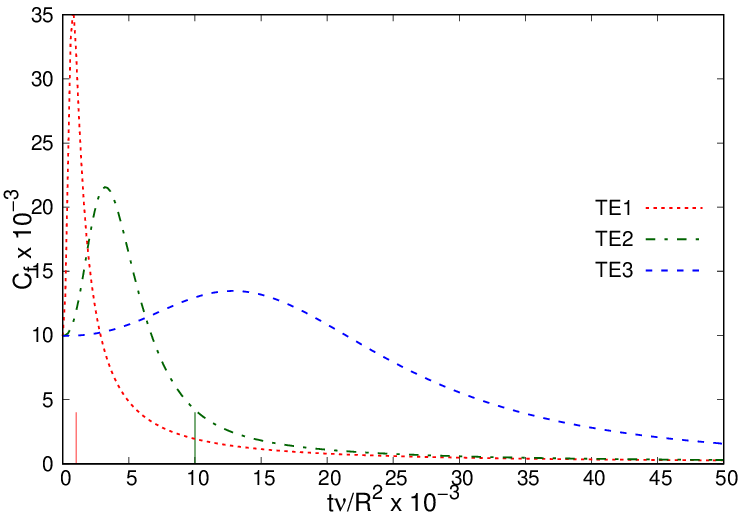}}
	\end{subfigure}
	\begin{subfigure}[\footnotesize{$C_f(\tau)$}]{\label{fig:CfLamT2}    \includegraphics[width=70mm]{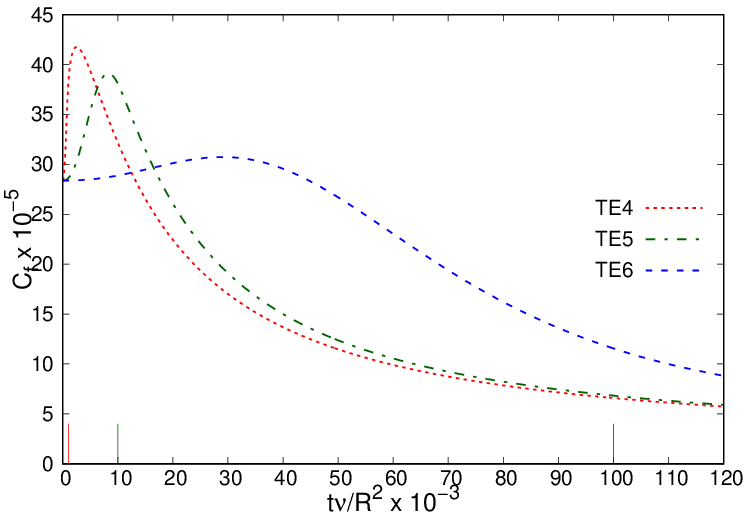}}
	\end{subfigure}
	\caption{\footnotesize{$Re(\tau)$, $|\sigma_w(\tau)|$, $C_f(Re)$ and $C_f(\tau)$ evolution for laminar U-flows TE1-TE6.}}
	\label{fig:TE1-TE6}
\end{figure}

Figs. \ref{fig:ReLam1}-\ref{fig:ReLam2} shows the evolution of ${Re}(\tau)$ for the theoretical experiments.
Note that, although the MPG increase is not a Heaviside step, the approximation $Re(\mathring{\tau}_c)\approx (1-\euler^{-1}){Re}_2$ is valid for TE1 and TE2 (Fig \ref{fig:ReLam1}), 
and for TE4 and TE5 (Fig \ref{fig:ReLam2}).
Thus, the resulting bulk U-flow for $\Delta \tau \ll \mathring{\tau}_c$ would not be noticeably different from the ideal canonical one caused by a Heaviside step
(which defines the time constant $\mathring{\tau}_c$), though locally they may differ. 
However, $\Delta \tau$ in TE3 and TE6 is comparable to $\mathring{\tau}_c$ and the ensuing U-flow is markedly different from the previous ones, at least during the beginning stage.
From about $\tau \approx 0.5$ onwards, all curves of each group become almost coincident among them, regardless of the thrusting interval $\Delta \tau$.
As a rule of thumb, it could be said that approximately for $\tau > 5 \Delta \tau$ the curve ${Re}(\tau)$ is almost indistinguishable from the ideal canonical one
(TE2 and TE5 are also almost indistinguishable from TE1 and TE4, respectively, for $\tau > 0.05$).
The rule can be tested by replacing $\tau$ by $5 \Delta \tau$ in Eqs. \eqref{eq:iniTransULF}, \eqref{eq:PresGradAfter} \& \eqref{eq:bulkSeries}.
The dynamics is fully characterised by the DoF $\Delta \Pi \mathop{=} \Pi_2-\Pi_1$ and $\Delta \tau$.

Figs. \ref{fig:WSSLam1}-\ref{fig:WSSLam2} show the evolution of absolute MWSS $|\sigma_w(\tau)|$ for both groups, TE1-TE3 and TE4-TE6, respectively.
Note how the initial increase is steeper than in Figs. \ref{fig:ReLam1}-\ref{fig:ReLam2}; actually the initial slope for $Re(\tau)$ is almost horizontal.
Thus, a fast growth of $C_f$ at the very beginning should be expected (see Eq. \eqref{eq:Cf}), which is exactly what the bathtub's initial sharp peak means.
The progression of $\sigma_w(\tau)$ is remarkably different from that of $\Pi(\tau)$, Fig. \ref{fig:RSSFT} \& Eq. \eqref{eq:AK+13PG}, 
being the fluid's mean bulk acceleration (MBA) the difference between them (recall $\Pi(\tau)$ only changes during $\Delta \tau$, otherwise it is constant).
It follows that the skin-friction coefficient $C_f$ (based upon $\sigma_w$) and the Darcy friction factor $f$ (defined through $\Pi$) 
cannot satisfy in U-flows the relation $C_f\mathop{=}f/4$, exclusive of S-flows.
Since $C_f \neq f/4$ in U-flows, an extreme caution must be exercised when attempting to use the Moody chart with any U-flow.
However, upon close examination of TE3 \& TE6 in Figs. \ref{fig:WSSLam1}-\ref{fig:WSSLam2}, during the initial instants of $\Delta \tau$ 
(at least up to $\Delta \tau/3$), the absolute MWSS follows rather faithfully the raising MPG, because the MBA is still relatively small.
Immediately after $\Delta \tau$, the MPG is constant and high ($\sim 1.45 \times 10^6$ in TE3 and $\sim 2.38 \times 10^6$ in TE6), the MWSS is still small ($<4 \times 10^5$ in all cases), 
and the difference is made up by the MBA.
This effect also occurs in TE1-TE2 and TE4-TE5, albeit it cannot be appreciated at the figures' scale.
Finally, much of what has just been said in the previous paragraph is also valid here: TE1 \& TE2 on one hand and TE3 \& TE4 on the other, are almost indistinguishable,
and for $\tau \mathop{\gtrsim} 5 \Delta \tau$ the curves are almost coincident with the canonical, regardless of $\Delta \tau$
(rule of thumb of $5\Delta \tau$).

Figs. \ref{fig:CfLam1}-\ref{fig:CfLamT2} show the evolution of transient $C_f$ for both groups TE1-TE3 and TE4-TE6.
Two sets of plots are displayed: $C_f({Re})$ in \ref{fig:CfLam1}-\ref{fig:CfLam2} and $C_f(\tau)$ in \ref{fig:CfLamT1}-\ref{fig:CfLamT2}.
Both depend on $\tau$, but their evolution rates are different. 
It is also plotted in Figs. \ref{fig:CfLam1}-\ref{fig:CfLam2} the hyperbola corresponding to laminar S-flow, $C_f\mathop{=}16/{Re}$, see \citep[Eq. (3.6)]{GF22}.
All curves of each group begin at the same point, corresponding to S-flow at $\tau\mathop{=}0$ of ${Re}_1\mathop{=}1600$, $C_f\mathop{=}16/1600\mathop{=}0.01$ for TE1-TE3 
and ${Re}_1\mathop{=}56223$, $C_f\mathop{=}0.0003$ for TE4-TE6.
We start with the ${Re}$-dependent curves of Figs. \ref{fig:CfLam1}-\ref{fig:CfLam2}.
We can see that initially the curves are overly removed from the steady-state $16/{Re}$ behaviour, to the point that $C_f$ increases with ${Re}$ instead of decreasing.
Then, after a maximum, sharper in TE1-TE3 than in TE4-TE6, the curves decrease rapidly approaching the steady-state $16/{Re}$ hyperbola, at which they tend asymptotically.
Short vertical lines in Figs. \ref{fig:CfLam1}-\ref{fig:CfLam2} mark the Reynolds number ${Re}(\Delta \tau_j)$ at which the increasing MPG becomes constant, being $j=1...6$ the index labelling TE1 to TE6. 
For $Re\mathop{>}{Re}(\Delta \tau_j)$ the U-flow moves under its own dynamics, for no increase of MPG is any longer occurring.
The calculated values of ${Re}(\Delta \tau_j)$ are shown in Table \ref{tab:laminarTE} (the short line corresponding to TE3 is not drawn,
because it falls outside the plot range).
The relative positions at which ${Re}(\Delta \tau_j)$ occur are, in increasing order: ${Re}(\Delta \tau_4)$ before the maximum of $C_f$ in TE4, 
${Re}(\Delta \tau_1)$ shortly after the maximum of TE1, ${Re}(\Delta \tau_5)$ also shortly after the maximum of TE5, 
${Re}(\Delta \tau_2)$ well within the decreasing stage when TE1 and TE2 almost coincide,
${Re}(\Delta \tau_6)$ when the three curves TE4-TE6 almost merge together (still far from the asymptotic zone of $16/{Re}$),
and ${Re}(\Delta \tau_3)$ already in the asymptotic zone of $16/{Re}$ (outside the plot range).
Only in TE4, most of the increasing behaviour of $C_f({Re})$ is still in progress after ${Re}(\Delta \tau_4)$, 
despite the MPG being constant, which demonstrates that the wall-shear stress is increasing faster than $Re^2$.
In TE1, $C_f$ already decreases sharply for ${Re}(\tau) \lesssim {Re}(\Delta \tau_1)$, despite the still increasing MPG, an effect explained by the raising MBA.
In TE1-TE3 the three curves merge together not far from the asymptote $16/{Re}$.
However, in TE4-TE6 the three curves $C_f({Re})$ converge amongst themselves much faster than with the S-flow $16/{Re}$ asymptote 
(convergence with the asymptote occurs beyond ${Re} \approx 5.0 \times 10^5$, not shown in the graph).

The width of the initial peak in $C_f({Re})$, stages i-ii of the bathtub, is approximately the same for TE1 \& TE2 on one hand,
and for TE4 \& TE5 on the other, since both groups verify $\Delta \tau \ll \mathring{\tau}_c$.
The width is smaller for TE3 and TE6, since $\Delta \tau$ is of the same order of magnitude as $\mathring{\tau}_c$.
The height of the initial peak, however, is different in each case, being very dependent upon $\Delta \tau$.
Also, the ${Re}$ at which the maximum $C_f$ is attained decreases with increasing $\Delta \tau$, albeit not much.
Otherwise put, the maximum occurs earlier as $\Delta \tau$ increases, which would seem a striking feature of this U-flow.
The TULF clearly explains and predicts this non-intuitive behaviour.

\begin{table}[h]
\centering
	\caption{\footnotesize{Data characterising theoretical experiments TE1-TE6 for pipe laminar flow.}}
	\label{tab:laminarTE}
	\footnotesize{
		\begin{tabular}{ccccccr}
			\toprule
			Id.  & ${Re}_1$ & ${Re}_2$ & $\Delta \tau$ & $\Pi_1$ & $\Pi_2$ & $Re(\Delta \tau_j)$\\
			\midrule
			TE1	& $1600$  & $3.6203 \times 10^5$ & $10^{-3}$ & $6.4 \times 10^3$	& $1.4481 \times 10^6$ &$ 2990$ \\
			TE2 & $1600$  & $3.6203 \times 10^5$ & $10^{-2}$ & $6.4 \times 10^3$	& $1.4481 \times 10^6$ & $14408$ \\
			TE3 & $1600$  & $3.6203 \times 10^5$ & $10^{-1}$ & $6.4 \times 10^3$	& $1.4481 \times 10^6$ & $98032$ \\
			TE4	& $56223$ & $5.9465 \times 10^5$  & $10^{-3}$ & $2.2489 \times 10^5$ & $2.3786 \times 10^6$ & $58299$ \\
			TE5 & $56223$ & $5.9465 \times 10^5$  & $10^{-2}$ & $2.2489 \times 10^5$ & $2.3786 \times 10^6$ & $75356$ \\
			TE6 & $56223$ & $5.9465 \times 10^5$  & $10^{-1}$ & $2.2489 \times 10^5$	& $2.3786 \times 10^6$ & $200280$ \\ 
			\bottomrule
		\end{tabular}
		}
\end{table}

The scenario just described becomes altered upon considering $C_f(\tau)$, Figs. \ref{fig:CfLamT1}-\ref{fig:CfLamT2}.
The width of the initial peak increases noticeably with $\Delta \tau$.
Also, the maxima occur later as $\Delta \tau$ increases, albeit the maxima are attained well before $\Delta \tau$, 
{i.e.}, $\tau_{max} <\Delta \tau$ for each experiment, except for TE4.
In TE4 $\Delta \tau$ is short and $\Delta \Pi$ is moderate, only one order of magnitude, while $\Delta \Pi$ is more than two orders of magnitude in TE1.
Furthermore, despite the higher excursion of $C_f(\tau)$, the curves reach the $16/{Re}$ asymptotic zone earlier for smaller $\Delta \tau$.
Thus, the lower $\Delta \tau$ the higher and sharper the peak, the more approximate $\tau_{max}$ and $\Delta \tau$ are,
and the sooner the asymptotic $16/{Re}$ is attained.
Great variations seem to be intense but short lived.

The experiments TE1-TE6 of laminar pipe flow have a faster evolution than similar experiments TE1-TE6 of channel flow, see \citep[Sect. 4]{GF24}.
The time range for curves of pipe flow is half the range of those for channel flow, despite the driving force (the MPG) being identical for both sets of experiments.
This is explained by the fact that the universal time constant for pipe flow, $\mathring{\tau}_c \approx 0.165$,
is some $2.4$ times smaller than $\mathring{\tau}_c \approx 0.4$ for channel flow. 
It follows that geometry matters, and when applying a fixed set of forces to various dynamical systems, their responses depend greatly upon the geometry. 
A first lesson learnt with laminar U-flows is that $C_f$ does not follow faithfully the MPG evolution (hence $C_f \neq f/4$).
Despite $\Delta \Pi \mathop{=} \Pi_2 - \Pi_1$ being identical for each set TE1-TE3 and TE4-TE6, $C_f$ appears to have its own dynamics.
A second lesson is that $C_f(\tau) \neq 16/{Re}(\tau)$ in laminar U-flow, although $C_f\mathop{=}16/{Re}$ is a correct result for S-flow.
A third lesson is that the initial peak of the bathtub of Fig. \ref{fig:bathtub} is inherited from the ULF, 
it is a typical and expected feature of laminar pipe flow.
Turbulence has nothing to do with it.
It follows that a theoretical and analytical explanation of the startling initial peak of the $C_f$ curve has been found; its dynamics is now disclosed.
Therefore, stages i, ii and iii of the bathtub, red in Fig. \ref{fig:bathtub}, 
are entirely and exclusively caused by the ULF.
Consequently, stages iv and v must be due to the turbulence (the PTC).
Finally, a fourth lesson is that the faster valve opening (smaller $\Delta \tau$), the higher Reynolds number at which the peak of $C_f$ is attained 
(see Figs. \ref{fig:CfLam1} and \ref{fig:CfLam2}), which constitutes a non-intuitive behaviour.

The key to understanding the curve of $C_f$ for laminar U-flow resides in the universal time constant $\mathring{\tau}_c \approx 0.165$, Eq. \eqref{eq:univCTE}.
It takes a relatively long time for the mean velocity to change in a laminar flow, due to its inertia.
However, if the MPG changes swiftly, then the MWSS must likewise change because the bulk velocity remains almost unchanged in such a short time.
Therefore, for small $\Delta \tau$, during the initial instants in $C_f\mathop{=}2 |\sigma_w | / \widetilde{u}^2$ ($\tau \lesssim \Delta \tau/3$),
the numerator $|\sigma_w|$ increases brusquely, whereas the denominator is largely unvarying.
It follows that an upward spike must be created in $C_f$, as observed in the figures.
But soon the MBA acquires significance, approximately for $\tau \gtrsim \Delta \tau/3$, and the MWSS must adjust to the freshly produced MBA,
and for $\tau > \Delta \tau$ the MPG becomes constant and only the MWSS remains to compensate for the changes in MBA.
With a significant MBA, the absolute MWSS increases slowly while $\widetilde{u}$ (and especially $\widetilde{u}^2$) does not cease to grow; 
in those conditions $C_f$ would normally decrease, and it keeps decreasing until the bulk velocity approaches its final steady-state value $\widetilde{u}_2$. 

This is all that laminar U-flows can do regarding $C_f$: nothing else may be expected from them.
Curves such as the bathtub must involve the turbulence and we shall learn in the coming section how the RSSRG shapes the mean flow to generate strange evolution curves for $C_f$.

\section{The skin-friction coefficient in the FTAM}
\label{sct:CfFrozenTurbulence}

In this section we shall study the conditions that lead to a $C_f(\tau)$ such as the bathtub, within the limitations of the FTAM, 
and it will be shown that not every U-flow yields a `\textit{canonical}' bathtub alike to Fig. \ref{fig:bathtub}.
Let be a frozen-turbulence U-flow starting in an initial S-flow with ${Re}_1$ and ending in another S-flow with ${Re}_2$, according to equations referred to in Apps. \ref{sct:modelS} \& \ref{sct:meanVelFieldS}.
Consider now an auxiliary U-flow, called the associated permanently-frozen-turbulence (PFT) U-flow, which has the following property:
the PFT U-flow begins with the same S-flow of ${Re}_1$ and is subject to the same MPG increase from $\Pi_1$ to $\Pi_2$, given by Eq. \eqref{eq:AK+13PG},
but the RSSRG remains stuck at that of $Re_1$ and does not evolve with time.
In fact, in a PFT U-flow only the ULF increases and evolves with time, while the PTC remains steady.
This PFT U-flow is considered canonical in a sense explained
below.
\begin{table}[h]
\centering
	\caption{\footnotesize{Initial and final states of transient pipe mean flows TEa to TEh in Table \ref{tab:TheorExp} (${Re}=2R\widetilde{U}/\nu$).}}
	\label{tab:FrozenTurbul}
	\footnotesize{
	\begin{tabular}{ccc|ccc}
		\toprule
		\multicolumn{3}{c|}{Flow definition} & \multicolumn{3}{c}{Pai's solution} \\
		\hline
		Id.  & ${Re}$ & $C_f$ & $\Pi$ & $q$ & $\chi$  \\
		\midrule
		PF1 &  $5300$ & $0.009193372$ & $6.45605 \times 10^4$ & $6$	& $4.62029$ \\
		PF2 & $14724$ & $0.006984237$ & $3.78539 \times 10^5$ & $14$	& $10.49621$  \\
		\bottomrule
	\end{tabular}
	}
\end{table}

\begin{figure}[h]
	\begin{center}
		\leavevmode
		\includegraphics[width=0.5 \textwidth, trim = 0mm 0mm 0mm 0mm, clip=true]{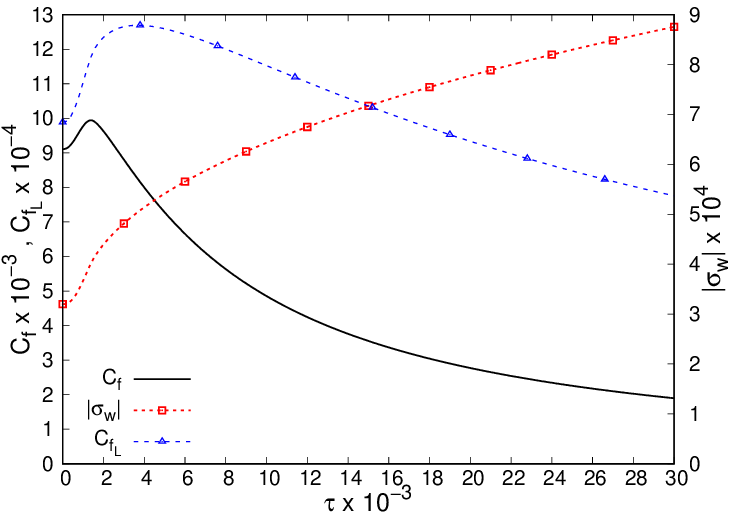}
		\caption{\footnotesize{Evolution of $C_f$, $C_{f_L}$ and $|\sigma_w|$ for the PFT U-flow ($\Delta \tau \mathop{=}0.0015$, $\mathring{\tau}_b\mathop{=}0.005458$).}}
		\label{fig:PipeFlowRefData}
	\end{center}
\end{figure}

Since to any conventional frozen-turbulence U-flow can be associated a PFT U-flow, let us consider the PFT U-flow that begins with PF1 of Table \ref{tab:FrozenTurbul},
and undergoes a MPG increase from $\Pi_1$ to $\Pi_2$ of PF2, but the RSSRG does not change and remains frozen at that of PF1.
This is analytically attained by assuming $\tau_0 \rightarrow \infty$ in Eq. \eqref{eq:RSSG}.
The ensuing PFT U-flow is not laminar, because it contains the relatively small PTC of PF1, but its laminarity increases with time as the ratio ULF/PTC.
Fig. \ref{fig:PipeFlowRefData} shows the evolution of $C_f(\tau)$ for such a PFT U-flow.
It also shows the ULF's skin-friction coefficient, $C_{f_L}(\tau)$, as reported in Sect. \ref{sct:laminarFriction}.
Here, $C_{f_L}$ is one order of magnitude lower, and also quite slower, than $C_f$.
It appears that the turbulence fosters the evolution of quantities in U-flows, even when the turbulence is steady (we shall see this is indeed true).
Finally, Fig. \ref{fig:PipeFlowRefData} also plots the absolute MWSS $|\sigma_w|$. It is observed that $|\sigma_w(\tau)|$ reproduces the evolution of $\Pi(\tau)$ during an interval $\sim \Delta \tau/3$,
because the bulk velocity increases very little during the initial stage of the U-flow.

Note $C_{f_2}\mathop{=}0.006984237$ for the final S-flow PF2, see Table \ref{tab:FrozenTurbul}.
Let $\mathring{\tau}_b$ be the earliest time at which $C_f(\mathring{\tau}_b)\mathop{=}C_{f_2}$, {i.e.}, 
the time at which the transient skin-friction coefficient of the PFT U-flow equals $C_{f_2}$.
Such $\mathring{\tau}_b$ is called the \textbf{canonical bathtub-time} and is defined for the canonical PFT U-flow; $\mathring{\tau}_b$ is needed for the classification of turbulent U-flows, as will be seen shortly.
In Fig. \ref{fig:PipeFlowRefData} it can be seen that $\mathring{\tau}_b\mathop{\approx}0.005458$.
Further, $\mathring{\tau}_b$ always occurs in stage iii of the
canonical bathtub.
Usually, it is $\mathring{\tau}_b > \Delta \tau$, unless $\Delta \tau$ is very large, 
which would imply a very slow opening of a valve.
The discussion of the canonical PFT U-flow ends here and we
return to the conventional frozen-turbulence U-flow, as defined in Sect. \ref{sct:model}.

It is also possible to define a bathtub-time $\tau_b$ for the conventional frozen-turbulence U-flow, 
analogous to the canonical $\mathring{\tau}_b$, through the equivalent condition $C_f(\tau_b)=C_{f_2}$.
Note that $\mathring{\tau}_b$ is meant for a canonical PFT U-flow, 
whereas $\tau_b$ is defined for a frozen-turbulence U-flow as per Sect. \ref{sct:model}. 
Typically, for an actual monotonously propelled mean flow, it would be $\tau_b\mathop{=}\mathring{\tau}_b$ and it would occur in stage iii of Fig. \ref{fig:bathtub}. 
However, there are cases in which $\tau_b\mathop{>}\mathring{\tau}_b$ and then $\tau_b$ occurs in stage v. 
With the nomenclature introduced in App. \ref{sct:modelS}, the frozen-turbulence U-flows may be classified as follows:
\begin{enumerate}[label=(\Roman*)]
 \item If $\tau_0\mathop{\leq}\mathring{\tau}_b$ the U-flow is of \textbf{early turbulence} 
 (and $\tau_b\mathop{>}\mathring{\tau}_b$, occurring in stage v).
 Most frequently, it would be $\Delta \tau \mathop{<} \tau_0 \mathop{\leq} \mathring{\tau}_b$, since there is frozen turbulence; see below.
 \item If $\tau_0 >\mathring{\tau}_b$ the U-flow is of \textbf{late turbulence} (in this case $\tau_b\mathop{=} \mathring{\tau}_b$ and it occurs in stage iii).
 \item If $\widehat{\Delta \tau}\mathop{=}\tau_2-\tau_0 \mathop{\leq} \Delta \tau$ the turbulence would be \textbf{fast}.
 \item If $\Delta \tau < \widehat{\Delta \tau}\mathop{\leq} 3 \Delta \tau$ the turbulence is deemed \textbf{moderate}.
\item If $\widehat{\Delta \tau} > 3\Delta \tau$ the turbulence is said to be \textbf{slow}.
\end{enumerate}
Determining how to experimentally achieve this catalogue of U-flows is beyond the scope of this research.
However, there are known methods of hindering or stimulating turbulence over a wide range.
There are four mean forces that configure the mean motion: MPG, MWSS, MBA (inertia) and RSSRG.
MPG only increases during $\Delta \tau$, being constant at any other time.
Likewise, RSSRG only changes during $\widehat{\Delta \tau}$.
Thereby, MWSS and MBA must change all the time to balance the remaining mean forces.
The reader is warned that the analysis to be developed in this section will focus primarily on forces, because only them can explain the shape of the bathtub (and the whole dynamics).

The TULF has the capability of completely separating the mean flows caused by different forces,
and follow their evolution in isolation, without the perturbing influence of other forces.
The structure of the general solution itself, Eq. \eqref{eq:equationGarcia}, shows clearly how and why is this possible. 
This sort of separate study is performed now, in the form of theoretical experiments with the U-flows of Table \ref{tab:TheorExp}.
All U-flows of Table \ref{tab:TheorExp} have identical initial and final S-flows, namely PF1 and PF2 of Table \ref{tab:FrozenTurbul}, 
which coincide very approximately with the initial and final S-flows reported in \citep{HSH16} and \citep[Table 1, case TP4]{GLC21} 
(note ${Re}=2\widetilde{U}R/\nu$ here, whereas ${Re}=\widetilde{U}R/\nu$ in those references).
The TDoF shown in Table \ref{tab:TheorExp} permit to distinguish one U-flow from any other; the remaining parameters are held identical in all U-flows.
Note the U-flows TEa-TEf have identical laminar component (the ULF).
TEg and TEh have faster ULF (smaller $\Delta \tau$), chosen to coincide with \citep{HSH16} and \citep[Table 1, case TP4]{GLC21}, although with identical initial and final S-flows.
Also, the levels of turbulence are identical in all experiments TEa-TEh; only the synchronisation of the turbulent component (the PTC) changes with respect to the laminar component (the ULF).
The main objective of these theoretical experiments is to understand why and
how different synchronisations yield entirely different friction profiles $C_f(\tau)$ in U-flows,
because synchronisation is all that changes from one to another.

The skin-friction coefficient $C_f$ of Table \ref{tab:FrozenTurbul} is calculated from the Darcy friction factor $f=4C_f$, 
which in turn is calculated from the Colebrook-White correlation 
\begin{equation}
 \label{eq:Colebrook}
 \frac{1}{\sqrt{f}} \mathop{=}-2 \ \log \left( \frac{\epsilon_r}{7.4R} + \frac{2.51}{{Re} \sqrt{f}} \right)
\end{equation}
where $\epsilon_r=0$ is the pipe roughness for smooth pipe.
Eq. \eqref{eq:Colebrook} is valid for ${Re}=2\widetilde{U}R/\nu$.
When calculating $C_f$, note that $\sigma_{w}\mathop{=}-\Pi/2$ for any S-flow.

\begin{figure}[]
\centering
	\begin{subfigure}[\footnotesize{TEa}]{\label{fig:TEa}
	    \includegraphics[width=70mm]{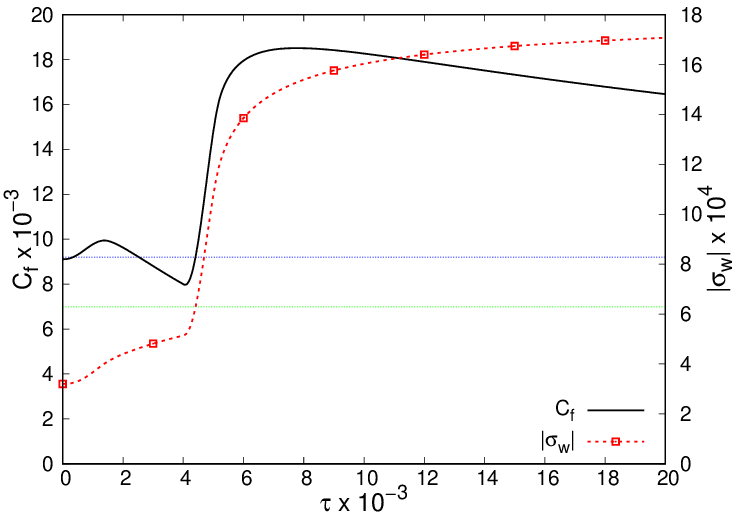}}
	\end{subfigure}
	\begin{subfigure}[\footnotesize{TEb}]{\label{fig:TEb}    \includegraphics[width=70mm]{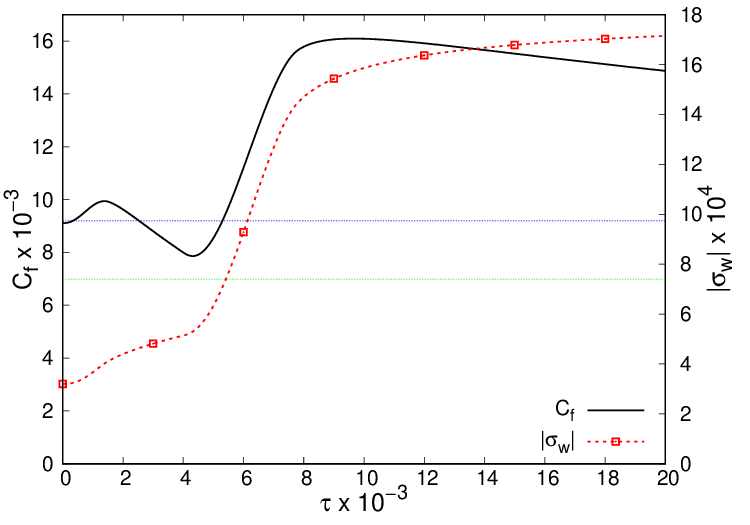}}
	\end{subfigure}
	\begin{subfigure}[\footnotesize{TEc}]{\label{fig:TEc}    \includegraphics[width=70mm]{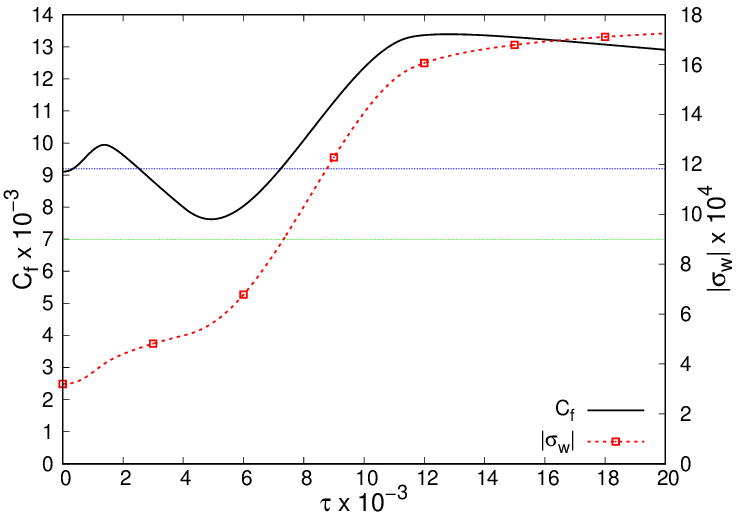}}
	\end{subfigure}	
	\begin{subfigure}[\footnotesize{TEd}]{\label{fig:TEd}    \includegraphics[width=70mm]{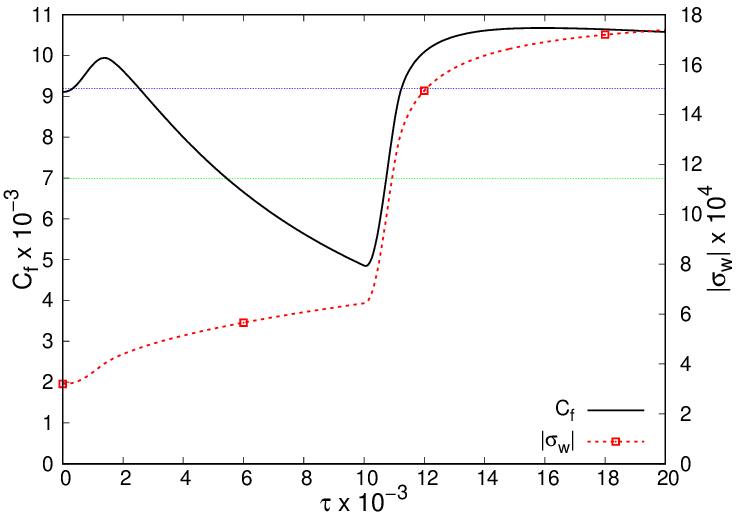}}
	\end{subfigure}
	\begin{subfigure}[\footnotesize{TEe}]{\label{fig:TEe}    \includegraphics[width=70mm]{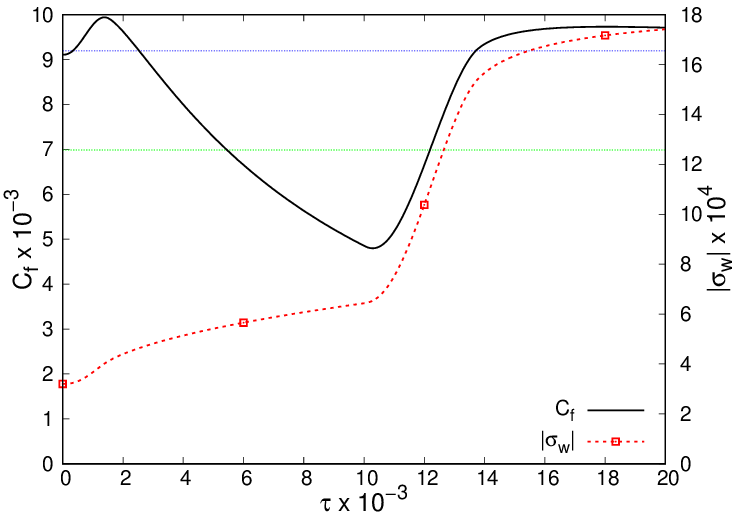}}
	\end{subfigure}	
	\begin{subfigure}[\footnotesize{TEf}]{\label{fig:TEf}    \includegraphics[width=70mm]{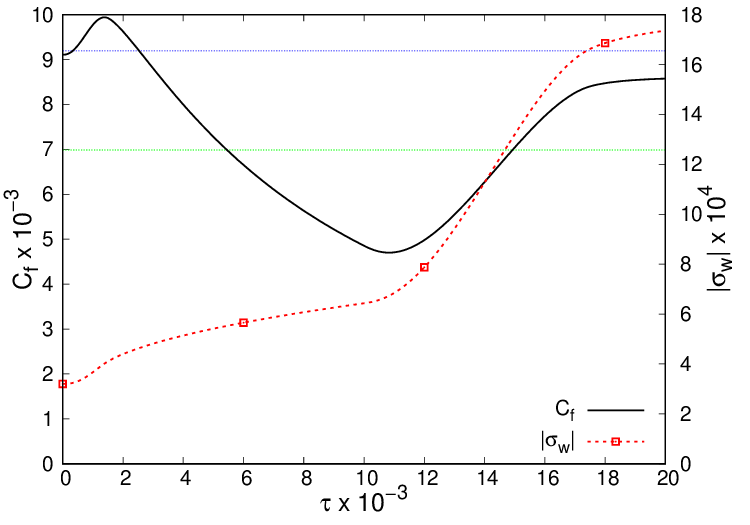}}
	\end{subfigure}	
	\begin{subfigure}[\footnotesize{Case TEg}]{\label{fig:TEg}
	    \includegraphics[width=70mm]{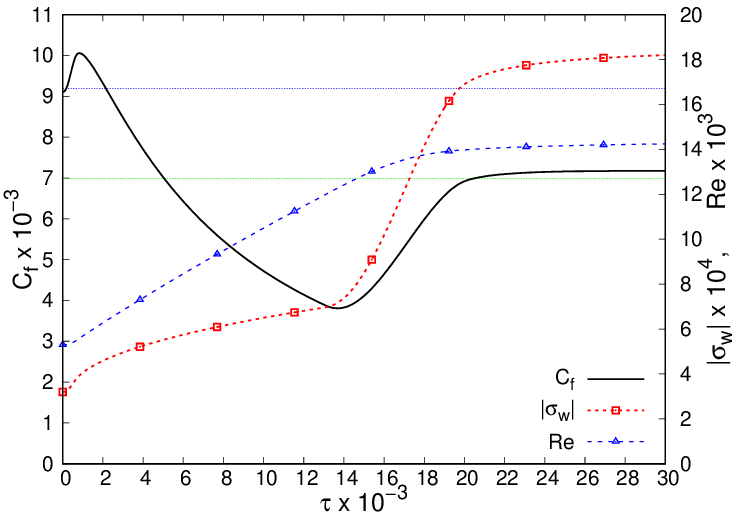}}
	\end{subfigure}
	\begin{subfigure}[\footnotesize{Case TEh}]{\label{fig:TEh} 
	    \includegraphics[width=70mm]{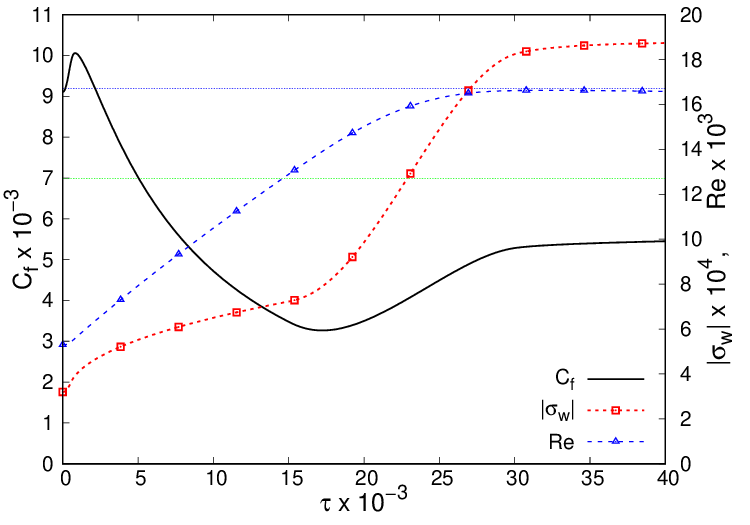}}
	\end{subfigure}
	\caption{\footnotesize{$C_f$ and $|\sigma_w|$ evolution for the FTAM. $Re$ is also shown in the `\textit{canonical}' bathtubs TEg \& TEh.}}
	\label{fig:TEa-TEh}
\end{figure}

In the experiments TEa-TEf of Table \ref{tab:TheorExp}, all corresponding curves in Fig. \ref{fig:TEa-TEh} will be coincident with Fig. \ref{fig:PipeFlowRefData} from  $\tau\mathop{=}0$ to $\tau\mathop{=}\tau_0$;
after $\tau_0$ they will begin to differ. 
Check also in Table \ref{tab:TheorExp} that early turbulence implies a very long bathtub-time ($\tau_b \mathop{\gg} \mathring{\tau}_b$),
well into stage v, whereas cases of late turbulence have their bathtub-time in stage iii ($\tau_b \mathop{=}\mathring{\tau}_b$).
Further, the values $C_{f_1}\mathop{=}0.009193372$ and $C_{f_2}\mathop{=}0.006984237$ of Table \ref{tab:FrozenTurbul} have been obtained from 
the Colebrook-White correlation, Eq. \eqref{eq:Colebrook},
whereas the values shown in Figs. \ref{fig:PipeFlowRefData}-\ref{fig:TEa-TEh} have been calculated from Eqs. \eqref{eq:bulkSeries}-\eqref{eq:WSSsymmet}.
It is reassuring to see such a coincidence in both sets of values, thus yielding additional proof of the formalism.

Finally, the influence of turbulence on the time constant of these U-flows will be evaluated.
Instead of considering the universal laminar time constant, $\mathring{\tau}_c \approx 0.165$, a modified version of this concept will be used:
$\tau_c$ is the time taken by the (non-laminar) bulk velocity $\widetilde{u}(\tau)$ to reach 
$\widetilde{u}(\tau_c)\mathop{=}(1-\euler^{-1}) \widetilde{u}_2$,
when perturbed with the actual forces given by Eqs. \eqref{eq:AK+13PG}, \eqref{eq:RSSG} and Table \ref{tab:FrozenTurbul}.
This new time constant $\tau_c$ differs from $\mathring{\tau}_c$ in two key aspects: there is a turbulent force (RSSRG),
and all applied forces follow a smooth evolution dictated by Eqs. \eqref{eq:AK+13PG}, \eqref{eq:RSSG}, instead of a Heaviside step.
The $\tau_c$ for TEa-TEh is shown in Table \ref{tab:TheorExp}.
In general, $\tau_c \ll \mathring{\tau}_c$, which means that turbulence greatly accelerates the evolution of mean U-flows.

The results of the theoretical experiments of Table \ref{tab:TheorExp} are displayed in Fig. \ref{fig:TEa-TEh}, and will be analysed in detail in the coming subsections.
The main objective of these theoretical experiments is to understand how and why different turbulence synchronisation yields entirely different friction patterns in pipe U-flow.
The study is developed within the limits of the FTAM.

\begin{table}[]
    \centering
    \caption{\footnotesize{Theoretical experiments in pipe U-flow ($\Delta \tau\mathop{=}0.0015$ for TEa-TEf; $\Delta \tau\mathop{=}0.0007$ in TEg-TEh).}}
    \label{tab:TheorExp}
    \footnotesize{
    \setlength{\arraycolsep}{2.25pt}
    \begin{tabularx}{\columnwidth}{@{}ccccccc|l@{}}
	\toprule
	Id.  & $\tau_0$ & $\tau_2$ & $\widehat{\Delta \tau}$ & $\tau_b$ & $\tau_c$ & $\tau_{80}$ & Type \\
	\midrule
	TEa	& $0.004$  & $0.0052$ & $0.0012\mathop{=}0.8 \Delta \tau$ & $0.599726$ & $0.0236578$ & --	& Early-fast turbulence\\
	TEb & $0.004$  & $0.00775$ & $0.00375\mathop{=}2.5 \Delta \tau$ & $0.5827824$	& $0.015106$ & -- & Early-moderate turbul.\\
	TEc & $0.004$  & $0.0115$ & $0.0075\mathop{=}5.0 \Delta \tau$ & $0.5540596$ & $0.00890$ & --	& Early-slow turbulence\\
	TEd	& $0.010$  & $0.0112$ & $0.0012\mathop{=}0.8 \Delta \tau$ & $0.005458$ & $0.00804216$ & $0.02868$ & Late-fast turbulence\\
	TEe & $0.010$  & $0.01375$ & $0.00375\mathop{=}2.5 \Delta \tau$ & $0.005458$ & $0.00804216$ & $0.01632$ & Late-moderate turbul.\\
	TEf & $0.010$  & $0.0175$ & $0.0075\mathop{=}5.0 \Delta \tau$ & $0.005458$ & $0.00804216$ & $0.01337$	& Late-slow turbulence\\ 
	TEg & $0.013$  & $0.02$ & $0.007\mathop{=}10\Delta\tau$ & $0.005077$ & $0.00765385$ & $0.01262$ & Very late-slow turbul.\\ 
	TEh & $0.015$  & $0.03$ & $0.015\mathop{=}21.4\Delta\tau$ & $0.005077$ & $0.00765385$ & $0.01262$ & Very late-slow turbul.\\ 
	\bottomrule
    \end{tabularx}
    }
\end{table}

\textbf{TEa}. 
The results of TEa, early and fast turbulence, are presented in Fig. \ref{fig:TEa}.
Two horizontal thin dotted lines mark the skin-friction coefficient of the initial and final S-flows, $C_{f_1}$ in blue and $C_{f_2}$ in green.
$C_f(\tau)$ is always above $C_{f_2}$.
It reaches $C_{f_2}$ at $\tau_b \mathop{=} 0.599726$ (see Table \ref{tab:TheorExp}), which is much outside the plotting range of Fig. \ref{fig:TEa}.
It is inherent to early turbulence that $C_f(\tau)$ reaches $C_{f_2}$ quite late, in stage v.
Thus, at any instant $\tau$ of this U-flow, $C_f(\tau)$ is much greater than the $C_f^S$ of corresponding equal-${Re}$ S-flow (see Sect. \ref{sct:Applications}),
and it would be erroneous to assume that a steady-state $C_f^S$ (from the Moody chart) may be used to calculate friction in these U-flows.
The maximum of $C_f(\tau)$, occurring at $\tau \approx 0.007$ in stage iv, is twice as large as the initial $C_{f_1}$.
The turbulence begins to unfreeze at $\tau\mathop{=}\tau_0\mathop{=}0.004$ and a steep increase is seen in $C_f(\tau)$ and $\sigma_w(\tau)$ from $\tau_0$ until $\tau \approx 0.006$,
roughly coincident with $\tau_2$.
The fact that both functions march almost parallel implies the bulk velocity is not changing much during the interval $(\tau_0,\tau_2)$.
However, Fig. \ref{fig:TEa} also shows that, for about $\tau \gtrsim 0.015$,
most variation in $C_f$ is due to the bulk velocity, since the MWSS is becoming steady.
It is also revealing that the increase in MPG during the interval $(0,\Delta \tau)\mathop{=}(0,0.0015)$ is much less important for $\sigma_w$ 
than the increase of RSSRG during $(\tau_0,\tau_0+\widehat{\Delta \tau})\mathop{=}(0.004,0.0052)$, 
an indication of which is more relevant to configure the friction.
Upon examining the evolution of absolute MWSS, $|\sigma_w(\tau)|$, and noting that the bulk velocity growth is relatively slow,
it would be understood why $C_f(\tau)$ undergoes the shown evolution.
Stages i to v can be distinguished in Fig. \ref{fig:TEa}, albeit stage iii is very short.

An important aspect that needs be highlighted is the influence of turbulence on the time constant of this mean flow,
considered as a dynamical system, which will be written $\tau'_c$ to distinguish it from the universal laminar time constant $\mathring{\tau}_c \approx 0.165$.
Such $\tau'_c$ would be canonically defined as the time taken by the (non-laminar) bulk velocity $\widetilde{u}(\tau)$ to attain 
$\widetilde{u}(\tau'_c)\mathop{=}(1-\euler^{-1}) \widetilde{u}_2\approx 0.63212\, \widetilde{u}_2$,
when perturbed with a unit Heaviside step in the forces (it must be stressed that $\tau'_c \not\approx 0.165$ in this case, for $\mathring{\tau}_c \approx 0.165$ only in laminar flow).
For practical purposes, however, a modified definition is used within this section: 
$\tau_c$ is the time taken by the (non-laminar) bulk velocity $\widetilde{u}(\tau)$ to attain 
$\widetilde{u}(\tau_c)\mathop{=}(1-\euler^{-1}) \widetilde{u}_2$,
when perturbed with the actual increase of forces given by Eqs. \eqref{eq:AK+13PG}, \eqref{eq:RSSG} and Table \ref{tab:FrozenTurbul}.
For TEa, such a time constant is $\tau_c\mathop{=}0.0236578$ and it decreases as advancing through the series from TEa to TEh, a phenomenon that is explained in Sect. \ref{sct:explTC}.
Note also that there is an order of magnitude between $\tau_c\mathop{=}0.0236578$ and $\tau_b\mathop{=}0.599726$, that is, 
the flow needs some twenty-five times longer to complete the remaining $\euler^{-1}\  \widetilde{u}_2$, to attain the final S-flow PF2.
This is typical behaviour of early turbulence.

It is quite interesting to observe in Fig. \ref{fig:TEa} how the MWSS approximately mimics what the RSSRG is doing (see Fig. \ref{fig:RSSFT}): 
when RSSRG is constant MWSS is (almost) constant and when RSSRG changes so does MWSS.
The MWSS reacts much quicker to the RSSRG than the PTC itself; otherwise put, the friction reflects the temporal changes in turbulence faster than the components of mean velocity.
In S-flow, it is not surprising that the shear stress follows closely the RSS: 
\citep[Eq. 3.3]{GF22} establishes the intimate relationship between $\dd u/\dd \alpha $ and $\sigma(\alpha)$.
An analogous (not identical) relationship exists for U-flow, see Eq. \eqref{eq:equationGarcia} and Eq. \eqref{eq:WSSsymmet}, 
in which emphasis is placed on RSSRG, rather than RSS.

\textbf{TEb}. 
Fig. \ref{fig:TEb} shows the results of TEb, early and moderate turbulence.
The vertical scale of $C_f(\tau)$ has shrunk respect to Fig. \ref{fig:TEa}, and it will keep decreasing as advancing through TEa to TEd,
meaning that the maximum $C_f(\tau)$ becomes smaller along the series.
On the contrary, the right-hand-side scale of MWSS remains identical in TEa-TEf.
The evolution of $C_f(\tau)$ for TEb is slower than for TEa, although not much.
However, the time constant for TEb, $\tau_c\mathop{=}0.015106$, is smaller than for TEa, which means a faster $\widetilde{u}(\tau)$ causing a slower $C_f(\tau)$.
The time constant $\tau_c$ will continue to decrease upon moving from TEa to TEf (see Sect. \ref{sct:explTC} for an explanation).
Also, stage iii of TEb's bathtub is still very short, albeit a bit longer than TEa's.
The duration of stage iii will increase as advancing through TEa to TEf.
Most observations made above for TEa are also applicable to TEb, and to the remaining theoretical experiments.

\textbf{TEc}. 
Fig. \ref{fig:TEc} shows the results of TEc, early and slow turbulence.
The increase of $C_f(\tau)$ after $\tau_0$ is not so steep in TEc and the left-hand-side scale has still reduced further
(a 70\% of TEa).
The growth of wall-friction is not so intense with slow turbulence, although $|\sigma_w(\tau)|$ reaches over $1.7\times 10^5$,
as in all other theoretical experiments.
The time constant for TEc is $\tau_c\mathop{=}0.00890$, noticeably shorter than those of TEa and TEb, and also stage iii has become longer than previous experiments.

\textbf{TEd}. 
Fig. \ref{fig:TEd} shows the results of TEd, late and fast turbulence.
This is the first instance where $C_f(\tau)<C_{f_2}=0.006984237$ for some time interval in stage iii; this event cannot occur with early turbulence,
as is obviously deduced from its definition. 
The sudden increase of $C_f$ and $|\sigma_w|$ after $\tau_0=0.01$ is again very steep, typical of fast turbulence.
It must be understood that temporal changes in turbulence, i.e. in RSS, are immediately converted into friction 
(compare $|\sigma_w|$ in Fig. \ref{fig:TEd} with RSS in Fig. \ref{fig:RSSFT}),
but they are not so swiftly transferred to the bulk velocity.
To grasp what turbulence does to MWSS, compare the curves of $|\sigma_w|$ in Figs. \ref{fig:PipeFlowRefData} and \ref{fig:TEd}:
at $\tau\mathop{=}0.02$ it is $|\sigma_w(0.02)|\lesssim 8 \times 10^4$ in the former and $|\sigma_w(0.02)|\gtrsim 1.7 \times 10^5$ in the latter, more than twice greater.
In turbulent U-flow, the initial increase of RSSRG is mostly converted into MWSS, and from it to $C_f$.
The increase of MPG during $\Delta \tau\mathop{=}0.0015$ is significantly less important for $|\sigma_w|$ as it is the increase of RSSRG during $\widehat{\Delta \tau}=\tau_2-\tau_0\mathop{=}0.0012$;
MPG causes one order of magnitude less effect in MWSS than RSSRG does.
The time constant for TEd is $\tau_c\mathop{=}0.00804216$, almost a third of TEa, which is also a fast turbulence U-flow.
Since $\tau_c<\tau_0$, it should be understood that $\tau_c$ be identical for TEd, TEe and TEf, because they all have the same $\tau_0$.
This is not a general characteristic of late turbulence; if $\tau_c>\tau_0$ then $\tau_c$ would decrease along the series TEd, TEe and TEf, as can be observed in \citep[Table 3]{GF24}. 
Since $\tau_c$ does not provide much information for TEd-TEf, it is replaced by $\tau_{80}$,
defined as the dimensionless time taken by the turbulent mean flow to attain $80\%$ of the final bulk velocity, $\widetilde{u}(\tau_{80})=0.8\ \widetilde{u}_2$. 
For TEd it is $\tau_{80}=0.02868$.
As a final remark, stage iii is becoming more noticeable in the bathtub of TEd, compared with the cases of early turbulence.

\textbf{TEe}. 
Fig. \ref{fig:TEe} shows the results of TEe, late and moderate turbulence.
Again, the skin-friction coefficient satisfies $C_f(\tau)<C_{f_2}$ for some time interval in stage iii, 
although the increase after $\tau_0$ is not so steep as in TEd.
The left-hand-side scale for $C_f(\tau)$ is still smaller in Fig. \ref{fig:TEe}.
It is observed in TEe that, for the first time in the series, the maximum value of $C_f(\tau)$ occurs at the initial peak, instead of in stage iv.
Such an effect could not be observed in the theoretical experiments developed for channel flow in \citep[Sect. 5]{GF24}.
For TEe it is $\tau_{80}= 0.01632$, shorter than for TEd.

\textbf{TEf}. 
Finally, Fig. \ref{fig:TEf} shows the results of TEf, late and slow turbulence.
The curve of $C_f(\tau)$ almost resembles a '\textit{canonical}' bathtub, though not quite yet.
The increase of $C_f(\tau)$ in stage iv is rather gentle.
For TEf it is $\tau_{80}=0.01337$, the shortest of all.
Since $\tau_{80}>\tau_0$ in TEd-TEf, it happens to be a suitable parameter to assess how fast these turbulent mean flows evolve.

In all theoretical experiments above, the increase of $C_f(\tau)$ in stage iv is caused by the MWSS $\sigma_w(\tau)$,
which shows a steep increase due to the unfreezing of turbulence during the time interval $(\tau_0,\tau_2)$.
Moreover, the arising of a new PTC also brings forth a decrease in the bulk velocity, albeit with some delay.
Of the two sources of motion, MPG and RSSRG (turbulence), the influence of the latter in MWSS is much greater than the former.
This phenomenon, and the timing at which it occurs, is possibly the most relevant result of the theoretical experiments.

We have seen the evolution of $C_f(\tau)$ for different synchronisations of turbulence in U-flows, TEa to TEf,
including a study of the time constant $\tau_c$.
Other values in Table \ref{tab:TheorExp} would have yielded different results;
the equations of App. \ref{sct:meanVelFieldS} dictate the actual shape of the curves shown in this research.
None of the resulting curves resembles yet the `\textit{canonical}' bathtubs reported in \citep[Fig. 2]{HSH16} and \citep[Fig. 2b]{GLC21},
from which follows that the dynamics governing the unfreezing of turbulence in those numerical experiments is not duly represented in the theoretical catalogue of TEa-TEf.
We shall see next what is needed to account for the actual experiments performed by \citep{HSH16} and \citep{GLC21}.

\subsection{Very late and very slow turbulence: cases \citep{HSH16} and \citep[TP4]{GLC21}.}
\label{sct:TEg}

With the knowledge gained in previous theoretical experiments, two skin-friction coefficient curves will be obtained 
with features that remember those shown in \citep[Fig. 2]{HSH16} and \citep[Fig. 2b, case TP4]{GLC21}.
Unfortunately, those papers did not report the MPG evolution in their DNS and, therefore, 
no information is available about one of the forces generating the mean motion of those U-flows.
Without knowing the forces, hardly can anyone determine the resulting movement (Newton's second law).
In Sect. \ref{sct:explBluntPeak} we shall offer a plausible explanation for this omission.
Thus, a time-dependent MPG ought to be proposed that yields curves alike to those reported, 
meaning that only the overall qualitative behaviour of $C_f(\tau)$ would be reproduced, rather than the exact simulations themselves.
A reasonable resemblance to \citep[Fig. 2]{HSH16} is found in Fig. \ref{fig:TEg}, herein called TEg.
This case is characterised by $C_f(\tau)$ being higher than $C_f$ corresponding to equal-$Re$ S-flow for $\tau \gtrsim \tau_2$ 
(see Sect. \ref{sct:Applications} and Fig. \ref{fig:TEg}).
In other words, $C_f(\tau)$ is above the green thin line marking $C_{f_2}$ for $\tau \gtrsim \tau_2$,
and this U-flow will be called overfrictioned.
Likewise, \citep[Fig. 2b, case TP4]{GLC21} is qualitatively emulated in Fig. \ref{fig:TEh}, called TEh,
and this case is characterised by $C_f(\tau)$ being for $\tau_{2}\lesssim  \tau\lesssim  \tau_{iv}$ lower than $C_f$ corresponding to equal-$Re$ S-flow (see again Sect. \ref{sct:Applications} and Fig. \ref{fig:TEh}),
or simply by noting that $C_f(\tau)$ is below the green thin line marking $C_{f_2}$ for $\tau \gtrsim \tau_2$,
and the U-flow will be called underfrictioned.
Those curves respond to the following parameters:
initial S-flow PF1, final S-flow PF2, $\Pi(\tau)$ given by Eq. \eqref{eq:AK+13PG}, 
$\varSigma(\tau,\alpha)$ by Eq. \eqref{eq:RSSG}, and the TDoF $\{$\textbf{TEg}; $\Delta \tau \mathop{=} 0.0007$,  
$\tau_0\mathop{=}0.013\mathop{=}18.6 \Delta \tau$, $\tau_2 \mathop{=} 0.02\mathop{=}28.6 \Delta \tau$, $\widehat{\Delta \tau}\mathop{=} 0.007\mathop{=}10\Delta \tau$, 
with $\tau_b\mathop{=} 0.005077$, $\tau_c\mathop{=}0.00765385$, $\tau_{80}\mathop{=}0.0126154$, the minimum of $C_f(\tau)$ occurring at $\tau_{iii}\mathop{=}0.0136923$ 
and $\tau_{iv}\mathop{=}0.029077 \}$ 
and $\{$\textbf{TEh}; $\Delta \tau \mathop{=} 0.0007$,  
$\tau_0\mathop{=}0.015\mathop{=}21.4 \Delta \tau$, $\tau_2 \mathop{=} 0.03\mathop{=}42.9 \Delta \tau$, $\widehat{\Delta \tau}\mathop{=} 0.015\mathop{=}21.4\Delta \tau \mathop{\gg} \Delta \tau$, 
with $\tau_b\mathop{=} 0.005077$, $\tau_c\mathop{=}0.00765385$, $\tau_{80}\mathop{=}0.0126154$ 
and the minimum of $C_f(\tau)$ occurring at $\tau_{iii}\mathop{=} 0.0172308 \}$. 
Both U-flows have the same $\tau_{80}$ because turbulence occurs quite late and the bulk velocity reaches $0.8\widetilde{u}_2$ before $\tau_0$, in each case.
This behaviour would not normally happen in plane-parallel U-flow, because it is much slower than pipe U-flow (see \citep{GF24}).
A consequence of underfriction, such as in TEh, is the rise of a velocity overshoot.
Note how $Re(\tau)>Re_2$ for $\tau>0.02$ in Fig. \ref{fig:TEh}.
This will be discussed in Sect. \ref{sct:explBathtub}.

Why the turbulence is actually so late and slow in the DNS of \citep{HSH16} and \citep{GLC21} is a question the TULF cannot answer,
for it is a theory of mean flows, not of individual physical flows (realisations, see \citep{GF25}).
It appears the RSSRG is only modified when the flow is not undergoing a significant MBA, i.e.,
meanwhile the MBA is high, the fluctuating velocity components would not engage in mutual correlations.
But, obviously, this behaviour belongs to the realm of physical flows and we may not adventure to explain it with our theory of mean flows.
It is observed in TEg and TEh, though, that $|\sigma_w(\tau)|$ increases significantly when RSSRG begins to increase ($\tau_0 \leq \tau \leq \tau_2$), 
and $\sigma_w(\tau)$ becomes almost steady when RSSRG ceases to change ($\tau>\tau_2$),
which implies that the MWSS follows quite faithfully the evolution of the bulk RSSRG.
To conclude, it is observed that a `\textit{canonical}' bathtub is only produced when the laminar TDoF (ULF) are much smaller than the turbulent TDoF (PTC),
namely $\Delta \tau \mathop{\ll} \tau_0$ and $\Delta \tau \mathop{\ll} \widehat{\Delta \tau} $;
otherwise, the resulting $C_f(\tau)$ curve would be alike to any of those of TEa-TEf.
This may be checked from the equations of App. \ref{sct:meanVelFieldS}.

\section{Discussion}
\label{sct:Discussions}

The reader should be aware that all results reported so far constitute predictions about the mean-flow fields and quantities arising in a U-flow, 
when it is subject to the initial conditions and forces, MPG and RSSRG, detailed in this work.
This connects with the direct problem of Mechanics, whereby the knowledge of applied forces grants the definition of the resulting movement. 
Above all, the reader must understand what it is and it is not being predicted. 
We do not know whether the actual forces acting on a pipe U-flow are as assumed here, but if they were, then the resulting mean flow must be as reported in these pages.
Therefore, if the mean-velocity field yielded by the TULF is anywhere near the experimental or simulation results,
it is because the MPG and RSSRG are also near to them, because the equation that governs the mean flow, the RANSE Eq. \eqref{eq:RANSEPP}, is linear.

We have seen in Fig. \ref{fig:TEa-TEh} that a `\textit{canonical}' bathtub is not always produced in a U-flow,
since it depends on the conditions in which the turbulence develops after being frozen, 
which must be very late and very slow ($\Delta \tau \ll \tau_0$ and $\Delta \tau \ll \widehat{\Delta \tau} $).
Upon reviewing the results of previous sections, it can be asserted about $C_f(\tau)$ that:
(I) the TDoF $\Delta \tau$ determines the height and width of the initial peak, stages i and ii of the bathtub 
(also the SDoF $\Delta \Pi \mathop{=}\Pi_2-\Pi_1$ influences the height of this peak); 
(II) the TDoF $\tau_0$ determines the position of the minimum in the bathtub, $\tau_{iii}$, with $\tau_{iii}>\tau_0$,
and (III) the TDoF $\tau_2$ (or rather $\widehat{\Delta \tau} \mathop{=}\tau_2-\tau_0$) determines the width of stage iv and the height of the second maximum,
which occurs at $\tau_{iv}> \tau_2$, because the PTC follows with some delay the evolution of RSSRG 
(the SDoF $\Delta\widehat{\Pi} \mathop{=} \widehat{\Pi}_2 - \widehat{\Pi}_1$ influences the height of this second maximum too).

The bathtub is the necessary consequence of the competition between the two components involved in any mean flow: the ULF and the PTC.
Their dynamical behaviour, the latter being delayed respect to the former, explains the peculiar shape of $C_f(\tau)$ during its temporal evolution.
However, the key question here is why the turbulence remains frozen for such a  long time. 
The PTC (which is the genuine measure of turbulence) does not change for a rather long time in TEg and TEh, 
and neither does in the DNS of \citep{HSH16} and \citep{GLC21}, which TEg and TEh emulate.
Why does the turbulence take $18.6\Delta \tau$ (TEg) or $21.4\Delta \tau$ (TEh) to begin to change?
Why does the RSSRG need $28.6\Delta \tau$ (TEg) or $42.9\Delta \tau$ (TEh) to become completely developed?
We think these are relevant questions, perhaps because the TULF has no answers for them.

It is now time for a detailed discussion about the bathtub itself.
More precisely, we need to discuss its stage iv, since stages i-iii have already been shown to correspond to laminar flow behaviour.
(see Figs. \ref{fig:CfLam1}, \ref{fig:CfLam2} and \ref{fig:PipeFlowRefData}),
and stage v is rather uninteresting because it shows the asymptotic descent to the final S-flow.
Stage iv is related to turbulence, albeit in a more surprising way than might initially be imagined.
The following issues will be discussed in this section:
\begin{enumerate}[label=(\Roman*)]
\item The interplay of forces leading to the bathtub, in particular its stage iv. 
The role of turbulence in shaping the mean-velocity field that causes the bathtub.
\item The deformation of mean-velocity U-profiles in the region that would have normally contained the log-law.
\item Why the time constant of the theoretical experiments TEa to TEh becomes increasingly shorter ($\tau_c$ or $\tau_{80}$).%
\item A detailed study of the unlikely physical reality of the U-flows reported in the DNS of \citep{HSH16} and \citep{GLC21}.
\end{enumerate}

The TULF deals with mean flows, which are not physical flows but mathematical entities not subject to the particularities of single realisations (see \citep{GF25}).
The predictions of the TULF are mathematical enunciates derived directly from the governing equation, which is linear.
All results presented in this research are the outcome of applying the mathematical equations shown herein and, thereby, they must be understood as mathematical theorems.

\subsection{On how the turbulence shapes the mean-velocity field to produce a bathtub}
\label{sct:explBathtub}

\begin{figure}[h!]
\centering
	\begin{subfigure}[\footnotesize{TEg: mean velocity $u(\tau,\alpha)$}]{\label{fig:TEg_Utime}
	    \includegraphics[width=70mm]{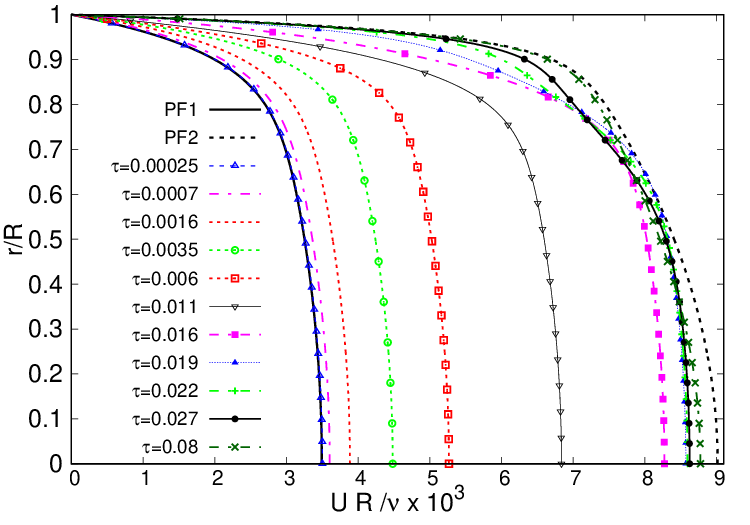}}
	\end{subfigure}
	\begin{subfigure}[\footnotesize{TEh: mean velocity $u(\tau,\alpha)$}]{\label{fig:TEh_Utime}    \includegraphics[width=70mm]{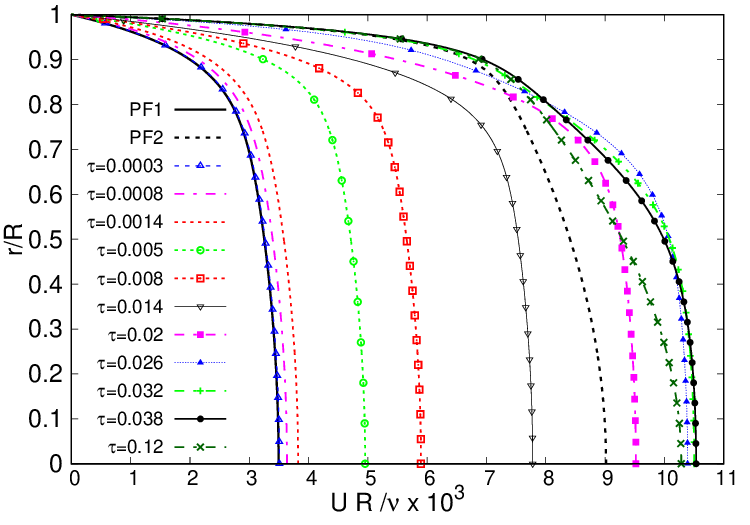}}
	\end{subfigure}
	\begin{subfigure}[\footnotesize{TEg: turbulence index $\mathfrak{I}(\tau,\alpha)={u(\tau,\alpha)}/{u_L(\tau,\alpha)}$}]{\label{fig:TEg_TItime}    \includegraphics[width=70mm]{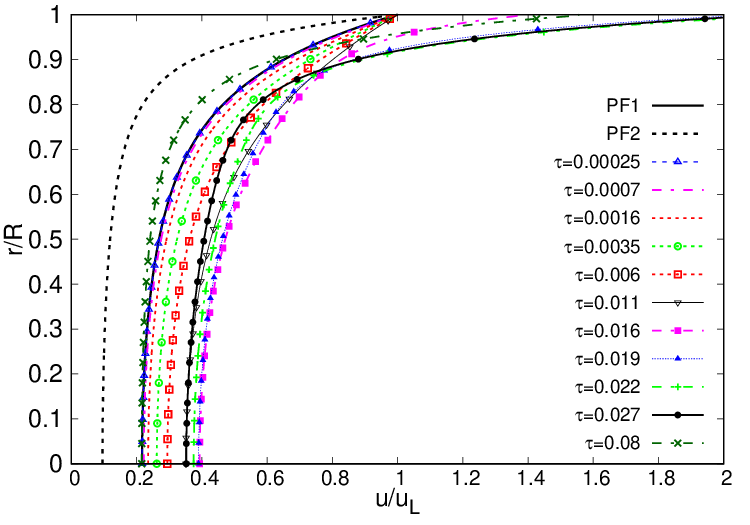}}
	\end{subfigure}	
	\begin{subfigure}[\footnotesize{TEh: turbulence index $\mathfrak{I}(\tau,\alpha)={u(\tau,\alpha)}/{u_L(\tau,\alpha)}$}]{\label{fig:TEh_TItime}    \includegraphics[width=70mm]{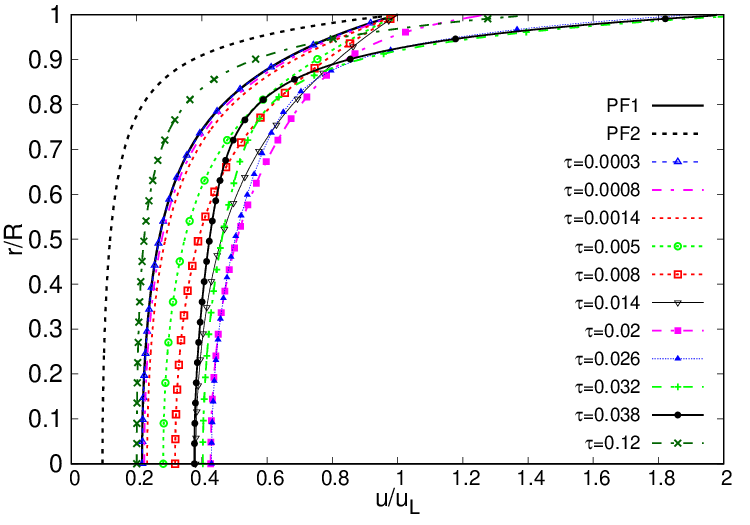}}
	\end{subfigure}
	\caption{\footnotesize{Time evolution of mean velocity and turbulence index fields for the FTAM. 
	The hyperlaminar sublayer (HSL) is the space-time domain where $\mathfrak{I}(\tau,\alpha)>1$.}}
	\label{fig:TEgTEhUTI}
\end{figure}

Arguably, the most effective approach to studying friction in U-flows is to examine its fundamentals. 
Therefore, we shall investigate the mean-velocity U-field that gives rise to friction, as this is where true knowledge lies.
The slope of the mean-velocity U-curve at the wall is particularly important, as it defines the MWSS.
Figs. \ref{fig:TEg_Utime}-\ref{fig:TEh_Utime} show the mean-velocity field $u(\tau, \alpha)$ for TEg and TEh, 
which illustrate the deformation undergone by mean-velocity U-profiles as different forces act during the evolution of U-flows.
This mean-velocity field has been calculated through Eqs. \eqref{eq:iniTrans}-\eqref{eq:RStressAfter}.
Mean-velocity U-profiles have been chosen for the instants: 
$\{$\textbf{TEg}: $\tau \mathop{=}$0.00025, 0.0007 (i), 0.0016 (ii), 0.0035, 0.006, 0.011 (iii), 0.016, 0.019, 0.022, 0.027 (iv) and 0.08 (v)$\}$
and $\{$\textbf{TEh}: $\tau \mathop{=}$0.0003, 0.0008 (i), 0.0014 (ii), 0.005, 0.008, 0.014 (iii), 0.02, 0.026 0.032, 0.038 (iv) and 0.12 (v)$\}$,
with at least one U-profile per each stage i to v of Fig. \ref{fig:bathtub} (stage iv is overpopulated because it is the most interesting).
Since the initial stages of TEg and TEh are identical up to the lowest $\tau_0$,
it has been deemed best to choose different instants in stages i, ii and iii for TEg and TEh,
so that the information reported in the figures be more extensive.

Upon comparing Figs. \ref{fig:TEg_Utime} and \ref{fig:TEh_Utime}, a neat feature stands out: 
all profiles in TEg are bounded between PF1 and PF2, whereas in TEh the mean-velocity profiles surpass PF2 for $\tau \geq 0.02$,
and for $\tau \gtrsim 0.026$ the profiles shrink back to PF2, i.e., there is a mean-velocity overshoot beyond PF2 (called \textit{accordion effect}, \citep{GF19c}).
The velocity in TEh exceeds that of TEg by up to $33\%$, despite the forces applied to both being equal in magnitude, 
though with different timing in the unfreezing of turbulence.
Such behaviour would not occur in the DNS of \citep{HSH16} and \citep{GLC21}, because ${Re}$ is held constant after the ramp.
But TEg and TEh are the result of applying relatively simple forces to the flow, and their $Re$ are not constrained to any constant value
and can evolve freely under their own dynamics.
With exactly the same MPG and the same level of RSSRG, by changing only the synchronisation of RSSRG,
the resulting bulk velocities of TEg and TEh differ remarkably. 
In this particular instance, it happens that TEg is an overfrictioned U-flow ($C_f(\tau)>C_{f_2}$ for $\tau>\tau_{iv}$, see Sect. \ref{sct:TEg}), 
whereas TEh is underfrictioned ($C_f(\tau)<C_{f_2}$ for $\tau>\tau_{iv}$), see Figs. \ref{fig:TEg}-\ref{fig:TEh}.
A U-flow with mean-velocity overshoots beyond PF2 is likely to be underfrictioned and a U-flow without overshoot is likely to be overfrictioned,
although it should not be considered a mathematical theorem, for it has not yet been proven in all cases.
This behaviour is also reported in \citep[Figs. 7 \& 8(a)]{GF24}, where a channel U-flow without mean-velocity overshoot is overfrictioned.

Zooming in near the wall in Fig. \ref{fig:TEg_Utime}, it can be checked that for $\tau<0.019$ 
the slope of the mean-velocity curve at the wall is clearly different from that of the final S-flow PF2.
On the contrary, for $\tau \geq 0.022$ the near-wall slope is indistinguishable from that of PF2;
that is, a negligible change is seen in the MWSS after the RSSRG becomes steady ($\tau_2=0.02$). 
The case $\tau\mathop{=}0.019$ shows a rather similar slope to PF2, although it is still somewhat different, as observed in the $\sigma_w$ curve of Fig. \ref{fig:TEg}.
For TEh, Fig. \ref{fig:TEh_Utime}, different wall slopes are found for $\tau \leq 0.02$, undistinguishable for $\tau \geq 0.032$ 
and slightly different slope to PF2 for $\tau=0.026$.
Therefore, despite the delay in the response of the PTC to changes in the RSSRG, as measured by its time constant, 
the MWSS follows quite faithfully the instantaneous evolution of the bulk RSSRG, with negligible delay.
This means that MWSS measurements can be used as indicators of RSSRG activity, a feature that experimental researchers would consider useful.

Note in Fig. \ref{fig:TEg_Utime} how for $\tau \gtrsim \tau_2$, {i.e.}, during and after the new turbulence becomes steady, 
most of the increase in mean velocity is created near the wall, for the core flow remains largely unchanged;
otherwise put, a significant growth of ${Re}$ in TEg occurs near the wall. 
We have the paradoxical result that the turbulence (dissipation) \textit{and} the mean velocity (new motion) are preferably generated at the same place. 
The core flow will begin to increase later, when the hyperlaminar
\footnote{A U-flow is said to be hyperlaminar at $(\tau,\alpha)$, \citep{GF24}, if there exists an open set $\Omega \subset (0,\infty)\times[0,1]$ containing $(\tau,\alpha)$, $(\tau,\alpha) \in \Omega$, 
such that $\mathfrak{I}(\tau',\alpha')>1$, $\forall (\tau',\alpha')\in \Omega$, being $\mathfrak{I}(\tau,\alpha)\mathop{=}u(\tau,\alpha)/u_L(\tau,\alpha)$ the turbulence index defined in \citep[Eq. 3.14]{GF22}.
Hyperlaminar U-flows tend to concentrate near the wall, where a \textit{hyperlaminar sublayer} (HSL) exists:
it is the spacetime domain where $\mathfrak{I}(\tau,\alpha)>1$ (see Figs. \ref{fig:TEg_TItime}-\ref{fig:TEh_TItime}).
The HSL has a time-variable thickness. There is no hyperlaminarity in S-flow.
Laminar U-flows satisfy $\mathfrak{I}(\tau,\alpha)\mathop{=}1$, everywhere and at all times.}
sublayer (HSL, see below) decreases its size;
for example, the U-profile at $\tau\mathop{=}0.08$ (stage v), much later after the RSSRG has become constant, 
shows only a modest increase in the core flow respect to the much earlier U-profile at $\tau\mathop{=}0.027$.
Despite being $\tau_c \ll 0.08$ for TEg, the core mean velocity has still a long way to go before reaching PF2.

The situation is different in TEh, Fig. \ref{fig:TEh_Utime}.
When the turbulence begins to unfreeze, $\tau_0\mathop{=}0.015$, the mean-velocity field is already almost as great as PF2.
The U-profiles for $\tau\mathop{\geq}0.02$ already have ${Re}(\tau)>{Re}_2$.
The growth in the core ceases about $\tau\mathop{=}0.026$ and, afterwards, the mean-velocity increases mostly near the wall.
The U-profile with maximum bulk velocity can be calculated doing $\dd \widetilde{u}/\dd t \mathop{=0}$ over Eq. \eqref{eq:bulkSeries}.
It occurs around $\tau \mathop{=}0.032$; thereafter the U-flow loses bulk velocity and approaches PF2.
Even the U-profile at $\tau \mathop{=}0.12$, which is much later than $\tau_2$, is still some $15\%$ greater than PF2.
The MPG pushes the U-flow, but the RSSRG pulls it; since both forces are not synchronised, the result is the accordion effect just described.

It follows that studying the bulk velocity, $\widetilde{u}(\tau)$, is insufficient for characterising accelerated U-flow (and its friction).
Instead, one must consider the entire mean-velocity field, $u(\tau,\alpha)$.
Furthermore, it is important to understand that the mean-velocity field continues to change over a relatively long period of time 
($\partial u(\tau,\alpha)/\partial \tau \neq 0$), despite the turbulence (and the MPG) having already become stationary 
($\partial \sigma(\tau,\alpha)/\partial \tau \mathop{=} \partial \varSigma(\tau,\alpha)/\partial \tau \mathop{=}0 $).

It is also observed in Fig. \ref{fig:TEg_Utime} that mean-velocity overshoots around $0.5 \lesssim \alpha \lesssim 0.8$ occur, after the new turbulence has entered into action.
To fix ideas, in a monotonously propelled mean flow a velocity overshoot occurs whenever and wherever $u(\tau',\alpha)<u(\tau,\alpha)$ for $\tau'>\tau$, 
i.e., the flow suffers a local deceleration despite being pushed ahead by active forces (the MPG).
Our theory predicts that overshoots will occur in places other than the wall and the centreline.
When the turbulence begins to unfreeze, there will be overshoots somewhere beyond the wall and somewhere out of the core flow.
This sort of behaviour should occur in U-flows leading to a `\textit{canonical}' overfrictioned bathtub like those of Figs. \ref{fig:bathtub} and \ref{fig:TEg}.
On the contrary, the overshoots in Fig. \ref{fig:TEh_Utime} occur everywhere for $\tau \mathop{\geq}0.02$ and, 
being ${Re}(\tau)$ greater than ${Re}_2$, it is only natural that $C_f(\tau)\mathop{<}C_{f_2}$, since the MWSS remains almost constant after $\tau_2$, see Fig. \ref{fig:TEh}.
Thus is explained the underfrictioned behaviour.

Furthermore, for $\tau \gtrsim 0.022$ the mean-velocity U-profiles show a slight concavity around $0.6 \lesssim \alpha \lesssim 0.9$ in Fig. \ref{fig:TEg_Utime}.
Readers will be familiar with mean-velocity profiles that are convex everywhere, and may be surprised to learn of locally concave profiles.
For example, the mean-velocity U-profiles of \citep[Figs. 2, 5, 8 and 11]{GF19c} are convex everywhere.
In certain cases, such concavity brings a change of sign in the slope of the mean-velocity tangent (the mean-velocity gradient).
This phenomenon was already reported in \citep[Fig. 15]{GF20}, where it received the name of \textit{lone concavity}, together with an analytical explanation of its occurrence.
Concavities in mean-velocity U-profiles have been reported in several experimental studies, most of which are referenced in our previous papers.
It is interesting to observe them in this research too, given that a sudden change in the mean-velocity gradient is an intrinsic property of certain U-flows.
Less obvious lone concavities are also present in Fig. \ref{fig:TEh_Utime}, particularly at $\tau\mathop{=}0.032$ 
and much less at $\tau\mathop{=}0.038$.

The physical explanation of concavities is surprisingly straightforward: the concavity exists because the near-wall U-flow moves faster than what the applied MPG would account for.
The MPG by itself can only cause everywhere-convex mean-velocity profiles.
This higher near-wall mean velocity is the empirical effect of hyperlaminarity, 
whereby a positive PTC supplies the additional mean-velocity observed in experiments and simulations.
The MPG, which causes the ULF, would not be able by itself to propel the fluid to such high mean velocities near the wall.
We have been taught that turbulence dissipates energy, but in the HSL the turbulence generates a measurable positive net kinetic energy, presumably extracted from the core flow.
The physical mechanism leading to such organised motion from turbulence remains unexplored, and can only be discovered through experimentation or DNS (realisations, see \citep{GF25}).

The evolution of the turbulence-index field $\mathfrak{I}(\tau, \alpha)$, shown in Figs. \ref{fig:TEg_TItime}-\ref{fig:TEh_TItime}, is no less interesting.
The turbulence index for PF2 is everywhere (except at the wall) smaller than for PF1, since PF2 is a more turbulent S-flow than PF1.
A HSL exists in TEg within the interval $0.915 \mathop{<}\alpha \mathop{<} 1.00$ for $ 0.016 \lesssim \tau < 0.08$,
being non-existent for $\tau\leq\tau_0$, because the turbulence is frozen and thus steady.
The maximum level of hyperlaminarity occurs at $\mathfrak{I}(0.022,0.99999)\mathop{=} 2.18$, outside the plotted range,
which is a remarkable hyperlaminarisation (the mean velocity is more than twice the ULF, $u\mathop{=}u_L+u_T \mathop{>} 2u_L$,  
which implies $u_T\mathop{>} u_L$, {i.e.}, the PTC is positive and greater than the ULF, instead of negative and smaller in absolute value).
Near the wall is the only place where it is guaranteed that ULF, PTC and mean velocity are all the same order of magnitude.
Anywhere else, typically ULF and PTC are same order of magnitude, but both much greater than mean velocity, 
which is the actual measurable quantity (see \citep[Fig. 2]{GF22}).
For TEh the hyperlaminarity ranges are similar to those of TEg: $0.925 \mathop{<}\alpha \mathop{<} 1.00$ for $ 0.026 \lesssim \tau < 0.12$.
The maximum hyperlaminarity occurs at $\mathfrak{I}(0.032,0.99999)\mathop{=}2.0576$, barely outside the plotted range.
The HSL in TEh lasts more than $0.12\mathop{-}0.015\mathop{=}0.105$, which is a rather long time.
Further, the HSL vanishes and becomes a laminar sublayer about $\tau\mathop{\approx}0.57$, 
with $\mathfrak{I}(0.57,0.99999)\mathop{=}1.001524\mathop{\approx}1$,
which means a duration of more than $0.555$ natural time units.
Undoubtedly, the HSL is a long-lasting structure of certain turbulent U-flows.

From the beginning of TEg and TEh, the turbulence index is greater than that of PF1, the initial S-flow.
All turbulence-index U-profiles before $\tau \mathop{\leq} \tau_0$ behave reasonably, being more laminar than PF1 (an expected feature), 
but without surpassing the supposedly limiting value of $\mathfrak{I}\mathop{=}1$.
Even, for $\tau \mathop{<} \Delta \tau$, the U-profiles of Figs. \ref{fig:TEg_TItime}-\ref{fig:TEh_TItime} are almost undistinguishable from PF1.
But as soon as the mean frozen-turbulence time $\tau_0$ is surpassed, $\tau_0\mathop{=}0.013$ in TEg and $\tau_0\mathop{=}0.015$ in TEh,
the hyperlaminarity appears.
All U-profiles of stage iv show a remarkable hyperlaminarity near the wall.
Even long after the RSSRG has become steady (for $\tau \mathop{\gg} \tau_2$ with $\partial \varSigma(\tau,\alpha)/\partial \tau \mathop{=}0$), 
the HSL still remains active, albeit it shrinks its thickness.
The stage-v U-profiles at $\tau\mathop{=}0.08$ in Fig. \ref{fig:TEg_TItime} and $\tau\mathop{=}0.12$ in Fig. \ref{fig:TEh_TItime},
are already more turbulent than PF1, except near the wall where they are still hyperlaminar.
Only the last U-profile of each series, $\tau\mathop{=}0.08$ for TEg and $\tau\mathop{=}0.12$ for TEh, is slightly more turbulent at the core than PF1; 
all others are everywhere less turbulent than PF1.
The U-flow's laminarisation endures even for $\tau \mathop{\gg} \tau_2$.

Recall hyperlaminarity is not possible without transient turbulence, {i.e.}, without a time-variable RSSRG,
although the HSL remains during a long time after the RSSRG has become steady, because so does the time-dependent positive PTC.
A monotonously accelerated laminar U-flow, however fast it might evolve, would only yield everywhere-convex mean-velocity U-profiles.
From this research follows mathematically that the `\textit{canonical}' bathtubs of Figs. \ref{fig:TEg}-\ref{fig:TEh} are not possible without hyperlaminarity,
and neither would the DNS bathtubs of \citep[Fig. 2]{HSH16} nor \citep[Fig. 2b, case TP4]{GLC21}.

Previous analyses have shown that the HSL constitutes a relatively stable and durable structure in the U-flow.
This determines how turbulence shapes the mean-velocity U-profile and drives turbulence diffusion throughout the pipe's cross section.
In particular, the HSL remains almost unaltered for at least the interval $0.019 \leq \tau \leq 0.027$ in TEg
and  $0.026 \leq \tau \leq 0.038$ in TEh.
Some of the energy drawn away from the core flow by the turbulence appears to be used to generate a new mean velocity within the HSL.
This mean velocity cannot be explained by active forces alone (MPG or gravity).
Unfortunately, the TULF provides no information about the dynamics of turbulence unfreezing, because this belongs to the realm of individual realisations (physical flows, see \citep{GF25}).
It can only be discovered through experimentation or simulation.

\subsection{The deformed log-law layer observed in U-flows}
\label{sct:linearNWVel}

\begin{figure}[h!]
\centering
	\begin{subfigure}[\footnotesize{Case TEg}]{\label{fig:TEg_Log} 
	    \includegraphics[width=70mm]{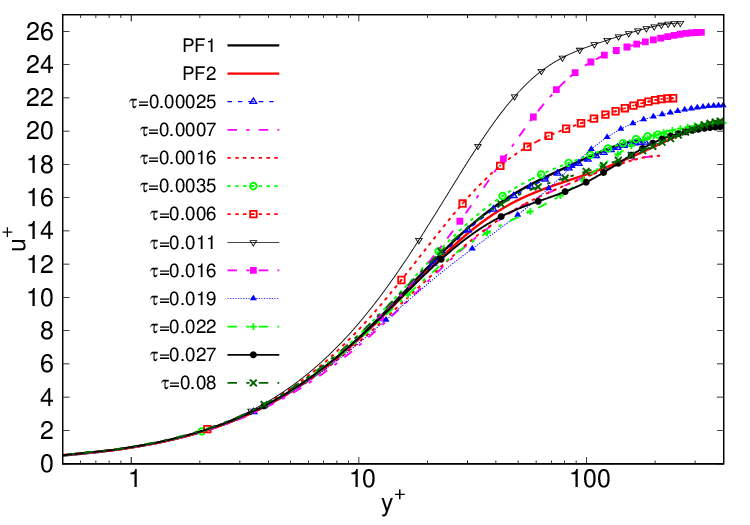}}
	\end{subfigure}
	\begin{subfigure}[\footnotesize{Case TEh}]{\label{fig:TEh_Log}
	    \includegraphics[width=70mm]{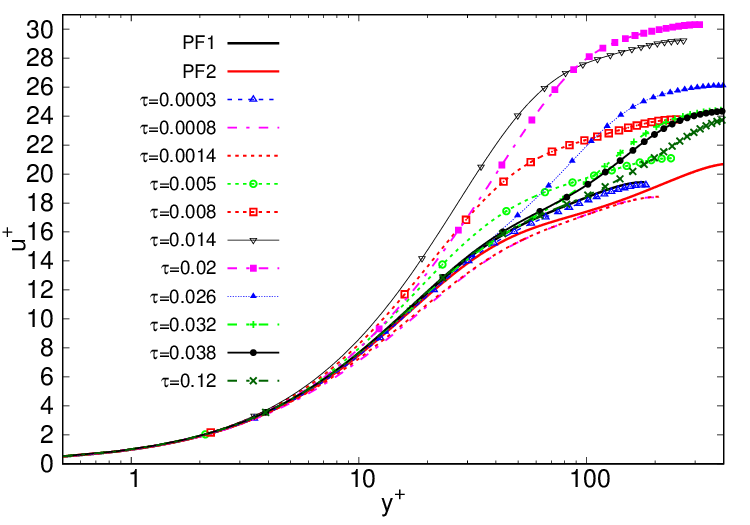}}
	\end{subfigure}
	\caption{\footnotesize{Time evolution of mean-velocity field for the FTAM (in wall units).}}
	\label{fig:TEgTEh_Log}
\end{figure}

The study of mean-velocity U-profiles shown in the previous section is enhanced upon considering the region normally occupied by the logarithmic layer.
This section refers to a type of U-flow description as reported in \citep[Fig. 7]{HSH16},
which explicitly shows the deformation of mean-velocity U-profiles expressed in wall units.
During the transient, the familiar structure of the logarithmic layer is destroyed.
The dimensionless wall distance and mean velocity are expressed in wall units as:
\begin{equation}
 \label{eq:wallUnits}
 y^+(\tau,\alpha) \mathop{=} (1-\alpha) u_{\tau}(\tau) \ , \qquad u^+(\tau,y^+)\mathop{=} \frac{u(\tau,\alpha)}{u_{\tau}(\tau)}
\end{equation}
whereby $y^+$ is also a time-dependent quantity and $u_{\tau}(\tau) \mathop{=} \sqrt{|\sigma_w(\tau)|}$ is the dimensionless friction velocity
(recall $R\mathop{=}\nu\mathop{=}\rho\mathop{=}1$ in the natural normalisation).
No attempt is made to express time in wall units, i.e., to define a $\tau^+$.
Figure \ref{fig:TEgTEh_Log} shows semi-logarithmic mean-velocity U-profiles for TEg and TEh, at same instants considered in Section \ref{sct:explBathtub}, expressed in wall units.
It also includes the familiar S-profile $u^+(y^+)$ corresponding to PF1 and PF2.
The structure of the logarithmic layer changes so much during the transient that it is no longer appropriate to call it a `\textit{logarithmic layer}'.
Even before the turbulence begins to unfreeze ($\tau\mathop{<}\tau_0$), mean-velocity U-profiles do not follow a logarithmic pattern.
The most deformed U-profiles occur shortly before and after $\tau_0$: $\tau\mathop{=}0.011$ and $\tau\mathop{=}0.016$ for TEg,
and $\tau\mathop{=}0.014$ and $\tau\mathop{=}0.02$ for TEh.
The unfreezing of turbulence affects differently the U-profiles, depending on whether the U-flow is over- or underfrictioned.
In overfrictioned TEg, the U-profiles at $\tau\mathop{\gtrsim}\tau_2$ are not so warped and especially the last two are rather close to PF2,
whereas in underfrictioned TEh none of shown U-profiles is close to PF2.

However, the near-wall linear structure $u^+(\tau)\mathop{=}y^+(\tau), \ y^+(\tau) \lesssim 5$, is conserved even in transient flow, 
although it should not be called `\textit{laminar sublayer}', because it may not be laminar, and the name `\textit{linear sublayer}' is adopted instead.
It was mathematically proved in \citep[Sect. 3]{GF22} that the viscous sublayer of S-flows is the physical manifestation of the ULF and, 
therefore, it is mean laminar, but this may no longer be true for U-flows, since Section \ref{sct:explBathtub} has shown a near-wall HSL.
The linear sublayer is a universal characteristics of Newtonian wall-bounded mean flows on which the no-slip condition be satisfied, and can also be observed in \citep[Fig. 7]{HSH16}.
The key issue here is not the specific values of $u^+(\tau,y^+)$, but rather the changing patterns adopted by the log-layer in U-flows.
It is the deformation exhibited by the transient curves when compared with the familiar logarithmic profile of an S-flow that is important.

\subsection{Explanation for the diminishing time constant}
\label{sct:explTC}

One among the many features that distinguish laminar from turbulent U-flows is the value of the time constant.
The '\textit{universal}' time constant of Eq. \eqref{eq:univCTE} holds only for laminar pipe U-flow; 
the time constant of turbulent U-flows is no longer universal and each U-flow endures its own.
The theoretical experiments of Sect. \ref{sct:CfFrozenTurbulence}, TEa to TEh, have shown that $\tau_c$  (or $\tau_{80}$ or any other value) for a mean flow decreases
as the turbulence becomes more delayed (see also Table \ref{tab:TheorExp}, columns $\tau_c$ and $\tau_{80}$), 
always provided that ${Re}(\tau_0) < (1-\euler^{-1}) {Re}_2$ because, otherwise, the time constant would reach its minimum value and would remain invariable.
The shortest $\tau_c$ of the series, $\tau_{c_g}\mathop{=}\tau_{c_h}\mathop{=}0.007654$, is over 21 times smaller than the universal time constant for laminar flow, $\mathring{\tau}_c \approx 0.165$.
Why does a turbulent flow react much faster than a laminar flow?
(here, '\textit{react}' means reaching a new steady bulk velocity after being perturbed stepwise).
The explanation is found in the mathematical realm of the TULF.
According to Table \ref{tab:FrozenTurbul}, in the initial S-flow, PF1, the ULF is some $\chi_1\mathop{=}u_{1_L}(0)/u_1(0)\mathop{=}\Pi_1/2u_1(0)\mathop{=}4.62029$ times greater than the mean velocity,
whereas in PF2 it is approximately $\chi_2\mathop{=}\Pi_2/2u_2(0)\mathop{=}10.49621$ times greater.
Thus, on average, the ULF of the U-flow from PF1 to PF2 would be $\tfrac 12 (4.62029+10.49621)\approx 7.6$ times greater than its mean velocity.
That is, $\widetilde{u}_L \approx 7.6\, \widetilde{u}$ and $\widetilde{u}_T \approx -6.6\, \widetilde{u}$ on average,
to yield $\widetilde{u}\mathop{=}\widetilde{u}_L+\widetilde{u}_T$ at all times (ULF and PTC).
It follows that the more delayed the turbulence, the more time the ULF would have to flow unopposed by the PTC,
and the faster the mean velocity will increase (recall, $\widetilde{u}\mathop{=}\widetilde{u}_L+\widetilde{u}_T$).
Every second the ULF is allowed to flow by itself, without the subtracting effect of the PTC, will make the mean velocity increasingly higher.
Therefore, the more delayed is the turbulence the faster will be reached the limit $(1-\euler^{-1}) \, \widetilde{u}_2 \approx 0.6321 \, \widetilde{u}_2$ 
and the smaller would be the time constant of the U-flow,
despite the fact that the time constant of a laminar U-flow (the ULF) is large.

In some cases, like TEh above, the turbulence could be so delayed that the bulk velocity would surpass the final $\widetilde{u}_2$, and a mean-velocity overshoot would occur.
This phenomenon is called the \textit{accordion effect} and it was first described and explained in \citep{GF19a},
and was first confronted with experimental evidence in \citep{GF19c}.
The behaviour of TEh, Fig. \ref{fig:TEh_Utime}, clearly reflects this surprising phenomenon.

\subsection{The unlikely evolution of ${Re}$ in relevant DNS simulations}\label{sct:explBluntPeak}

It was already remarked in Sect. \ref{sct:TEg} that \citep{HSH16} and \citep{GLC21} did not report the forces applied to the flow in their DNS, and how unusual seemed that omission,
since the absence of forces brings those research works into the realm of Kinematics, rather than Dynamics.
Here it will be shown that the forces needed to cause the reported U-flows are so extraordinary, that they may render the U-flows physically unrealisable.
Otherwise put, it would be very unlikely to ever witness the reported flow-rates (${Re}(\tau)$) in actual laboratory experiments.

Those DNS constitute examples of the so-called inverse problem of Mechanics, 
whereby the function that describes motion (output) is prescribed and the forces that would generate such motion (inputs) ought to be calculated.
In terms of Newtonian Mechanics, $\mathbf{F}=m\ddot{\mathbf{x}}$, it would be equivalent to prescribe $\mathbf{x}(t)$ 
(or $\dot{\mathbf{x}}(t)$) and purport to determine the $\mathbf{F}(t)$ causing it.
While the inverse problem generally has mathematical sense, its solution may be physically unlikely if the prescribed motion is not sufficiently regular
or if the governing equation is non-linear or both.
However, well-posed direct problems always yield sound physical solutions, even if the only available solving method is a computational simulation.
In principle, it is perfectly possible to feed a time-dependent pressure-gradient boundary condition to a DNS and collect the motion data it produces,
which will be experimentally reproducible as long as the input pressure gradient is plausible.

We purport to reproduce the ${Re}(\tau)$ curves of \citep[Fig. 1]{HS13} (which is replicated in \citep{HSH16}) and \citep[Fig. 2a]{GLC21} using
direct methods,
i.e., prescribing a MPG and obtaining those ${Re}(\tau)$.
In fact, only the initial ramp needs be reproduced, for the remaining flow occurs at constant $Re(\tau)=Re_2$.
Furthermore, one should only consider the laminar component, the ULF, to reproduce the initial ramp, since the turbulence is frozen and no new PTC is generated during such a short interval.
Thus, let us assume the S-ULF of PF1, subject to MPG $\Pi_1$ and with Reynolds number $Re_{L_1}=\Pi_1/4$ (see the Zero Theorem in \citep[Eq. 3.6]{GF22}),
and suppose that at $\tau=0$ it undergoes a Heaviside step in MPG up to $\Pi_{\infty}$, such that at $\Delta \tau^*$ the U-ULF attains the Reynolds number of PF2, $Re_{L_2}=\Pi_2/4$,
being $\Delta \tau^*$ the duration of the ramp.
The values of $\Pi_1, \, Re_1$ and $\Pi_2, \, Re_2$ are found in Table \ref{tab:FrozenTurbul}.

To calculate the ULF, one only needs to consider Eq. \eqref{eq:equationGarcia} without the PTC term.
The initial velocity field is a Hagen-Poiseuille S-flow:
\begin{equation}
 \label{eq:Hagen-Poiseuille}
 u_{L_0}(\alpha)= \frac{\Pi_1}{4}(1-\alpha^2)=\sqrt{2}\ \Pi_1 \sum \limits_{n=1}^{\infty}\frac{\phi_n(\alpha)}{\lambda_n^3}
\end{equation}
and the U-ULF takes the form:
\begin{align}
 \label{eq:U-ULF}
 &u_L(\tau,\alpha)=\sqrt{2}\ \Pi_1 \sum \limits_{n=1}^{\infty}\frac{\euler^{-\lambda_n^2 \tau}\phi_n(\alpha)}{\lambda_n^3} + \sqrt{2}\sum \limits_{n=1}^{\infty}\frac{\phi_n(\alpha)}{\lambda_n} \int \limits_0^{\tau}\Pi(\tau') \euler^{-\lambda_n^2 (\tau-\tau')} \dd \tau'= \notag \\
 &\sqrt{2}\sum \limits_{n=1}^{\infty} \left( \frac{\Pi_1 \euler^{-\lambda_n^2 \tau}}{\lambda_n^3}+ \frac{\Pi_{\infty} (1-\euler^{-\lambda_n^2 \tau})}{\lambda_n^3} \right) \phi_n(\alpha)=
  \frac{\Pi_{\infty}}{4}(1-\alpha^2) - \sqrt{2} (\Pi_{\infty}-\Pi_1) \sum \limits_{n=1}^{\infty}  \frac{ \euler^{-\lambda_n^2 \tau} \phi_n(\alpha)}{\lambda_n^3}
\end{align}
The bulk velocity corresponding to this field is given by Eq. \eqref{eq:bulkSeries}, resulting in:
%
\begin{equation}
 \label{eq:U-bulkULF}
 \widetilde{u}_L(\tau)=\frac{\Pi_{\infty}}{8} - 4 (\Pi_{\infty}-\Pi_1) \sum \limits_{n=1}^{\infty}  \frac{ \euler^{-\lambda_n^2 \tau} }{\lambda_n^4}
\end{equation}
At $\tau\mathop{=}\Delta \tau^*$ the ULF's Reynolds number ${Re}_L(\tau)$ should equal ${Re}_L(\Delta \tau^*)\mathop{=}{Re}_{L_2}\mathop{=}\Pi_2/4$, and
\begin{equation}
 \label{eq:fin_bulkULF}
 \Pi_2 =\Pi_{\infty}- 32 (\Pi_{\infty}-\Pi_1) \sum \limits_{n=1}^{\infty}  \frac{ \euler^{-\lambda_n^2 \Delta \tau^*} }{\lambda_n^4}
\end{equation}
Everything is known in Eq. \eqref{eq:fin_bulkULF} except $\Pi_{\infty}$, whose value is:
\begin{equation}
 \label{eq:PiInfinite}
 \Pi_{\infty} = \left( \Pi_2 - 32 \Pi_1  \sum \limits_{n=1}^{\infty}  \frac{ \euler^{-\lambda_n^2 \Delta \tau^*} }{\lambda_n^4} \right ) \left[ 1-32 \sum \limits_{n=1}^{\infty}  \frac{ \euler^{-\lambda_n^2 \Delta \tau^*} }{\lambda_n^4} \right]^{-1}
\end{equation}
For water ($\nu\mathop{=}\SI{1.0533e-6} {\metre \squared \per \second}$ at $\SI{18}{ \degreeCelsius}$) and pipe radius ${R}\mathop{=}\SI{25}{\milli \metre}$, 
the interval $\Delta t^*\mathop{=}0.22$ reported in \citep[page 132]{HSH16} equals $\Delta t\mathop{=} \SI{0.0177}{\second}$.
Such short interval equals $\Delta \tau^* \mathop{=} 3\times 10^{-5}$ in the natural normalisation. 
Inserting it into the series of Eq. \eqref{eq:PiInfinite} and calculating with 250 terms, the series yields:
%
\begin{equation}
 \label{eq:serieL4}
\sum \limits_{n=1}^{\infty} \frac{\euler^{-\lambda_n^2 \Delta\tau^*}}{\lambda_n^4} = 0.031242561691279317
\end{equation}
and the requested pressure gradient takes the value $\Pi_{\infty}\mathop{=}1319158523.04518\mathop{\approx}1.32\times 10^9$,
which  in a Hagen-Poiseuille flow would correspond to a Reynolds number ${Re}_{L_{\infty}}\mathop{=}\Pi_{\infty}/4\mathop{=}3.3\times 10^8$,
i.e., a dimensional ULF bulk velocity in a DN50 pipe of $\widetilde{U}_{L_{\infty}} = \SI{6947.35}{\metre \per \second}$
(recall dimensionless mean velocities are designated with lowercase $u$ and dimensional with uppercase $U$).
The result deserves to be repeated, should it not be clear enough: to reproduce experimentally the DNS of \citep{HSH16}, it would be necessary to apply during $\Delta t\mathop{=} \SI{0.0177}{\second}$ a MPG which,
if not deactivated, would finally push the ULF to the unlikely bulk velocity of $\widetilde{U}_{L_{\infty}}\mathop{=}\SI{6947.35}{\metre \per \second}$ in a DN50 pipe full of water.
After a long time, when the turbulence unfreezes and the mean flow becomes steady, such an ULF would correspond to a turbulent mean flow of ${Re}_{\infty}\mathop{\approx}1.4 \times 10^6$, see \citep[Eq. (7.4)]{GF22}, 
which in a DN50 pipe would yield a measurable mean bulk velocity of  $\widetilde{U}_{{\infty}}\mathop{\approx}\SI{29.3}{\metre \per \second}$.
Of course, it is impossible to apply a Heaviside step in laboratory and the actual required MPG $\Pi_{\infty}$ would have to be higher, possibly even more than double.
The other quantities would have to increase accordingly.

The U-flow reported in \citep{HSH16} would be obtained after deactivating $\Pi_{\infty}$ at $\Delta \tau^*$ and replacing it with $\Pi_2$,
the final MPG.
The bulk velocities $\widetilde{U}_{L_{\infty}}$ and $\widetilde{U}_{{\infty}}$ would only occur if $\Pi_{\infty}$ is not deactivated.
How could two so different velocities as $\widetilde{U}_{L_{\infty}}$ and $\widetilde{U}_{{\infty}}$ be related,  
if they are separated by more than two orders of magnitude?
The link between them is established through a huge accordion effect, much greater than the theoretical example reported in \citep[Fig. 3b]{GF19a}.
At all times, it is $\widetilde{U}\mathop{=}\widetilde{U}_L\mathop{+}\widetilde{U}_T$, with $\widetilde{U}_L\mathop{\geq}0$ and $\widetilde{U}_T\mathop{\leq}0$;
but $\widetilde{U}_T$ is frozen at a small initial value and does not change during a long interval,
while $\widetilde{U}_L$ increases enormously until reaching $\widetilde{U}_{L_{\infty}}$.
Therefore, the mean bulk velocity $\widetilde{U}$, which is the actually measurable quantity, increases initially only with $\widetilde{U}_L$ and should reach extremely high values.
Later, when the turbulence unfreezes, the PTC will also increase very much in absolute value, decreasing the mean bulk velocity up to the final value of $\widetilde{U}_{{\infty}}$.
This just described up-and-down behaviour of the mean bulk velocity $\widetilde{U}$ constitutes the accordion effect.

The numbers for \citep[case TP4]{GLC21} are more moderate, but still difficult to achieve in practice.
In their nomenclature, for a MPG Heaviside step in water and DN50 pipe, those numbers are: $\Delta t^{+0}_{ramp} \mathop{=}23.5$, $\Delta  t_{ramp} \mathop{=} \SI{0.477}{\second}$, 
$\Delta \tau^* \mathop{=}8 \times 10^{-4}$, the series of Eq. \eqref{eq:serieL4} equals $0.031058430422759319$, $\Pi_{\infty}\mathop{=}5.13 \times 10^7$, 
${Re}_{L_{\infty}}\mathop{=}\Pi_{\infty}/4\mathop{=}1.3\times 10^7$ and $\widetilde{U}_{L_{\infty}}\mathop{=}\SI{270.1}{\metre \per \second}$.
Whenever the turbulence unfreezes and the mean flow becomes steady, it would be: ${Re}_{\infty}\mathop{\approx} 2.325 \times 10^5$ 
and $\widetilde{U}_{{\infty}}\mathop{\approx}\SI{4.9}{\metre \per \second}$,
also some two orders of magnitude lower than $\widetilde{U}_{L_{\infty}}$.
For a physical non-Heaviside step, these numbers should be increased accordingly.

\section{Applications}
\label{sct:Applications}
This section discusses some practical applications of the findings uncovered in this research, 
and it follows quite faithfully the line of argumentation initiated in \citep[Sup-Sect. 7]{GF24}.
We shall see in which circumstances is the friction in a U-flow lower or higher than in an equivalent S-flow.
It is assumed that the reader is already familiar with the notion of SB-flow at $\tau$ and
its skin-friction coefficient $C_{f_{\tau}}^S$ associated to $C_f(\tau)$, introduced in \citep[Sup-Sect. 7]{GF24}.
Here $C_{f_{\tau}}^S$ is calculated through the Colebrook-White correlation, Eq. \eqref{eq:Colebrook}.

\begin{figure}[h]
\centering
	\begin{subfigure}[\small{Case TEg}]{\label{fig:TEg_Cf} 
	    \includegraphics[width=70mm]{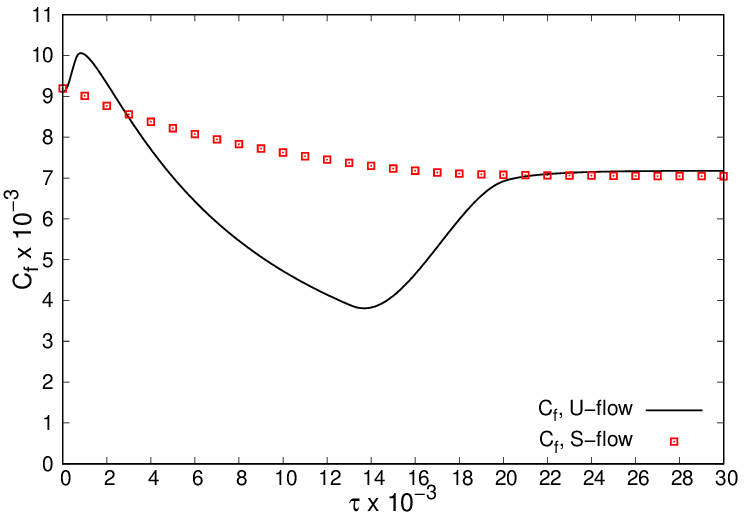}}
	\end{subfigure}
	\begin{subfigure}[\small{Case TEh}]{\label{fig:TEh_Cf}
	    \includegraphics[width=70mm]{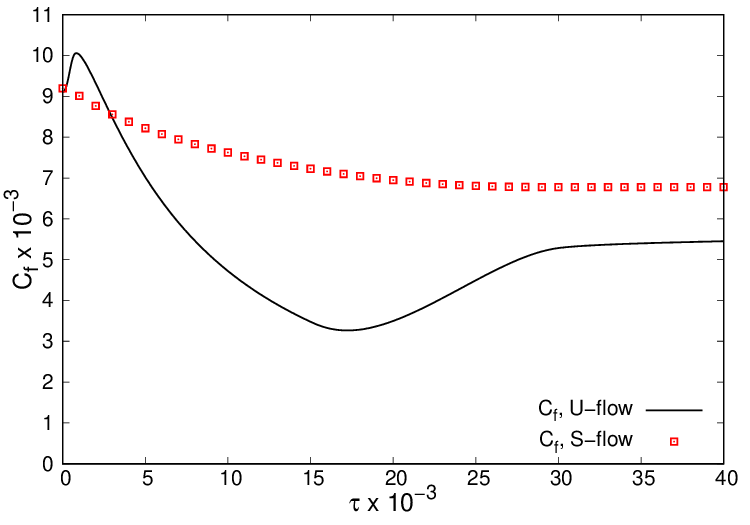}}
	\end{subfigure}
	\caption{\small{Comparison of $C_f$ for U-flow and equal-${Re}$ S-flows (SB-flows).}}
	\label{fig:TEgTEh_Cf}
\end{figure}

Since both, the U-flow and SB-flows, have the same ${Re}$ at each $\tau$, their skin-friction coefficients $C_f(\tau)$ and $C_{f_{\tau}}^S$
would be a measure of the relative magnitude of their respective MWSS.
Figure \ref{fig:TEgTEh_Cf} shows the bathtubs of TEg and TEh and the corresponding $C_{f_{\tau}}^S$,
for instants evenly spaced in 0.001 natural time units.
Note the different behaviour of U-curves in stages iv and v, which would justify the names of `over-' and `underfrictioned' U-flows.
In a U-flow like TEh, which endures an accordion effect, the skin-friction coefficient is always lower than corresponding SB-flows,
except during the initial peak.
In a U-flow like TEg, $C_f(\tau)$ is only lower than corresponding $C_{f_{\tau}}^S$ in stages iii and iv.
Practitioner engineers should take note of these possibilities in their applications with transient flows.

The situation in non `\textit{canonical}' U-flows, such as TEa to TEf above, would be different and 
the U-flow would present a higher skin-friction than corresponding SB-flows.
Depending on whether one needs high or low friction, a sensible choice of a turbulence development pattern
(early, late, fast, moderate, slow and combinations thereof) will help in achieving the objectives of most applications.
Current technology permits the selection of turbulence enhancers or inhibitors that would lead a flow to adopt anyone of the patterns discussed above, 
and thus take advantage of the specific properties of each class of U-flow for practical applications. 
This research offers relevant information for determining how and when to profit from the intervals of higher or lower friction in the U-flow.

An interesting application, also discussed in \citep{GF24}, would be the Clean-in-Place (CIP) technique for cleaning the insides of pipes without dismounting them.
By carefully generating a suitable transient flow, the MWSS of stages i and ii in the `\textit{canonical}' bathtub (or stage iv in earlier turbulence) 
would be sufficiently intense to remove most dirt adhered to a wetted surface.
The calculation procedure shown in Sect. \ref{sct:explBluntPeak} would yield quite approximately the necessary MPG $\Pi_{\infty}$
to produce a desired MWSS in stages i and ii.
The practitioner engineer should select suitable MWSS and peak duration $\tau_{ii}\mathop{=}\Delta \tau^*$ to remove dirt,
and then apply equations offered in Sect. \ref{sct:explBluntPeak} to design the U-flow.
The U-ULF's MWSS is obtained from Eq. \eqref{eq:WSSsymmet}, with the following result:
\begin{equation}
 \label{eq:U-MWSS}
 \sigma_{w_L}(\tau)=-\frac{\Pi_{\infty}}{2} + 2 (\Pi_{\infty}-\Pi_1) \sum \limits_{n=1}^{\infty}  \frac{ \euler^{-\lambda_n^2 \tau} }{\lambda_n^2}
\end{equation}
Since the MWSS of initial S-flow is given exclusively by the ULF (\citep[see][Sect. 3]{GF22}), 
and since only new ULF is generated during stages i and ii (because the PTC is frozen), 
it follows that Eq. \eqref{eq:U-MWSS} would be appropriate to calculate the actual U-flow's MWSS.

\

\section{Conclusions}
\label{sct:Conclusions}

The present research has developed a comprehensive study of the friction in monotonously propelled pipe mean flow, 
considering its causes and the timing of concurrent forces that generate the motion.
Even uncommon aspects such as the decrease in the time constant with turbulence delay, or the influence of hyperlaminarity in increasing friction, have been analysed.
Also, it has been shown that the mean-velocity field continues to change for relatively long after the RSSRG has become stationary and, in some cases, like underfrictioned U-flows, for quite longer.
This work provides a new insight into how the development of turbulence affects the friction in U-flows.
This insight is not based on the degree of turbulence, but on its timing,
since all studied cases reach the same turbulent S-flow at the end.
Experimental researchers are encouraged to explore this largely uncharted pool of phenomena, since this paper contains an unusually high number of predictions.
The theory can be used to guide the design of experiments.

The main conclusions of this study can be summarised as follows: 

(I) The equations of the mean-velocity field for fully-developed pipe U-flow according to the FTAM have been first published in App. \ref{sct:meanVelFieldS}.

(II) In laminar U-flow, $C_f(\tau)$ has an initial peak and then decreases smoothly and asymptotically to the $16/{Re}$ behaviour of laminar S-flow.
No other evolution can occur in a monotonously accelerated laminar U-flow.

(III) Most of the initial motion of a monotonously propelled turbulent mean U-flow is explained by the exclusive action of its ULF, 
since the PTC remains frozen in its initial value.
Although $C_{f_L}(\tau)$ in laminar U-flow is typically one order of magnitude lower than $C_f(\tau)$ in frozen-turbulence U-flow, they share an analogous initial behaviour.

(IV) The evolution of friction in U-flows depends enormously on the delay in creating new turbulence.
If the new turbulence delays too much, the resulting U-flow will develop an accordion effect and it will be underfrictioned, as in TEh, 
being its transient friction lower than that of corresponding equal-$Re$ SB-flows most of the time (see Sect. \ref{sct:Applications}).
If the new turbulence is sufficiently slow, but not too much, as in TEg, the U-flow will generate a `\textit{canonical}' bathtub 
and the transient friction would be lower than in corresponding SB-flows during stages iii and iv, but then it will become overfrictioned in stage v.
With early (or not so late) turbulence the transient friction remains very high during most of the U-flow, higher than in corresponding SB-flows.

(V) The time constant of the U-flow decreases as the turbulence becomes more delayed, and it is at least one order of magnitude lower than the universal time constant $\mathring{\tau}_c$.

(VI) The MWSS appears to follow rather faithfully the evolution of the bulk RSSRG in a U-flow, with negligible delay, although such a general statement must still be rigorously proved.
It may be of interest for experimental researchers.

(VII) Hyperlaminarity must occur in a U-flow that presents a `\textit{canonical}' bathtub, causing a local mean-velocity overshoot (herein called a lone concavity).

(VIII) The U-flow that produces a `\textit{canonical}' bathtub has a relatively stable and long-lasting HSL.
The HSL persists long after the RSSRG has become steady.
This behaviour occurs regardless of whether the U-flow is over- or underfrictioned.
Such an HSL must be investigated experimentally, and the community is encouraged to discover the mechanisms by which turbulence creates a new mean velocity within this sublayer
(the MPG, gravity and other active forces cannot create it).

(IX) The log-law layer is destroyed during the U-flow, but recovers its familiar shape in stage v, some time later.

(X) The layer of U-flow that touches the wall, defined by $u^+(\tau) \approx y^+(\tau), \ y^+(\tau) \lesssim 5$, is called the linear sublayer.
The linear sublayer is a universal property of Newtonian mean flows in contact with walls (no-slip condition).
The linear sublayer of a S-flow is mean laminar and, actually, it is the ULF.

(XI) Using a relatively simple model (see Sect. \ref{sct:model}), the TULF has predicted the main results reported in \citep{HSH16} and \citep{GLC21}.
Otherwise put, the TULF provides a causal explanation of the reported facts, based on the general principle that forces cause motion.

This is the eighth of a series of articles that will explain a number of uncommon phenomena already reported in the literature,
for which an analytical explanation within the framework of the traditional theory does not seem to be available.
An analogous study of friction as that reported herein is offered in \citep{GF24}, but for channel flow. 
\\

\noindent  \textbf{Data availability}\quad\footnotesize   The data that support the findings of this study can be readily calculated from the equations published on it.
All reported results were obtained using a laptop computer with i7 microprocessor, 16 GB RAM and standard Python libraries.   
\\

\noindent  \textbf{Author contributions and ORCID}\label{sct:Contributions}\quad\footnotesize   FJGG (0000-0002-8065-1449): conceptualization, investigation, methodology, mathematical development, software writing, data gathering, figures, manuscript writing (initial draft) and editing.

PFA (0000-0002-9598-5249): supervision, critical revision, validation, editing and funding acquisition.   


\section*{Declaration of interests}
The authors report no conflict of interest.

\bibliographystyle{alpha}
\bibliography{J123_02R00_Bathtub_Bibliography}

\section*{Appendices}

\appendix

\section{The frozen-turbulence analytical model (FTAM)}
\label{sct:modelS}

This section presents the three mathematical functions that must be prescribed to uniquely determine the mean-velocity field $u(\tau,\alpha)$ defining any U-flow,
according to the GAS of Eq. \eqref{eq:equationGarcia}:
the initial mean-velocity field $u_0(\alpha)$, the MPG $\Pi(\tau)$ and the RSSRG $\varSigma(\tau,\alpha)$.

\subsection{Analytical model for the initial mean-velocity field}
\label{sct:Uzero}

First, the initial mean-velocity field $u_0(\alpha)$ of the U-flow needs be modelled.
It is assumed that the U-flow begins as a S-flow of ${Re}_1$, whose general expression in terms of RSS, in natural units (see Sect. \ref{sct:background}), is given by\citep{GF20}
\begin{equation}
 \label{eq:PipeSflowU}
 u_1(\alpha)=\frac{\Pi_1}{4}(1-\alpha^2)\mathop{-}\int\limits_{\alpha}^1 \sigma_1(\alpha')\dd\alpha'\mathop{=} \frac{\Pi_1}{4}(\frac{1}{\chi_1}-\alpha^2)\mathop{+}\int\limits_{0}^{\alpha} \sigma_1(\alpha')\dd\alpha'
\end{equation}
The mean-velocity S-field is suitably modelled with a Pai's polynomial like in \citep[Eq. 7.1]{GF22}:
\begin{equation}
\label{eq:PaiFlow}
u_0(\alpha)\mathop{\equiv}u_1(\alpha)\mathop{=} \frac{\Pi_1 }{4 \chi_1} \left( 1+ \frac{\chi_1-q_1}{q_1-1}\ \alpha^2 -\frac{\chi_1-1}{q_1-1}\ \alpha^{2q_1}\right)\mathop{=} 
\frac{\Pi_1}{4}(1-\alpha^2)+ \frac{\Pi_1(\chi_1-1)}{4\chi_1(q_1-1)} \left[ 1 - q_1\left(1-\alpha^2\right)-\alpha^{2q_1} \right]
\end{equation}
whereby $\Pi_1$ is the MPG, $q_1 \in \mathbb{N}$, $q_1\mathop{\geq}2$ a best-fitting integer and $\chi_1=u_{1_L}(0)/u_1(0) \in \mathbb{R}$, $1\mathop{\leq} \chi_1 \mathop{<}q_1$,
the so-called {Centreline Turbulent Dissipation} (CTD).
This set of parameters is called the Spatial Degrees of Freedom (SDoF) of the initial S-flow.
The first expression in Eq. \eqref{eq:PaiFlow} is called the \textbf{compact form} of Pai's flow,
whereas the second is the \textbf{decomposed form}, for it shows explicitly the ULF
($\Pi_1 (1-\alpha^2)/4$) and the PTC (the remainder).
For moderate ${Re}$, this polynomial matches very well the Princeton Superpipe data shown in \citep[Fig. 8]{GF22}.
The final S-flow is also described by an identical polynomial with SDoF $\Pi_2$, $q_2$ and $\chi_2$.

\subsection{Analytical model for the mean-pressure gradient (MPG)}\label{sct:AMMPG}


The MPG source function needs also be modelled.
It is assumed that a constant MPG $\Pi_1$ drives the initial S-flow.
Then, at $\tau\mathop{=}0$, a valve is opened and the MPG begins to increase until it reaches $\Pi_2$, where the MPG becomes stationary.
The mean valve-aperture time required to drive the transient U-flow is modelled with the parameter $\Delta \tau$,
and constitutes the sole DoF required for $\Pi(\tau)$; actually it is a temporal DoF (TDoF).
Once the valve is fully open, no further variation is expected in the MPG,
even though the flow has still a long time to reach the final steady-state ${Re}_2$ (longer than the time constant of Eq. \eqref{eq:univCTE}).

The proposed time-dependent MPG is given by:
\begin{align}
\label{eq:AK+13PG}
\Pi(\tau)\mathop{=}\begin{cases}
\Pi_1 \ \ &if \ \ \tau <0 \\
\Pi_1 \left( 1+2\left(\frac{\tau}{\Delta \tau}\right)^2 \left( \frac{\tau}{\Delta \tau}-\frac{3}{2}\right) \right) -2 \Pi_2 \left( \frac{\tau}{\Delta \tau} \right)^2 \left(\frac{\tau}{\Delta \tau}- \frac{3}{2} \right)  \ \ &if \ \ 0 \leq \tau \leq \Delta \tau \\
\Pi_2 \ \ &if \ \ \tau > \Delta \tau \\
\end{cases}
\end{align}
with $\Pi_1, \Pi_2 \in \mathbb{R}$ the initial and final constant MPG, respectively, driving the initial and final S-flows.
Mathematically, this evolution is a nonlinear homotopy between the initial and final values, $\Pi_1$ and $\Pi_2$, with continuous derivatives, 
and the homotopy parameter is $\tau/\Delta \tau$, running from $0$ to $1$.
In plain English: the MPG changes monotonously from $\Pi_1$ to $\Pi_2$ and yields an accelerated mean flow, 
with $\dd \Pi/\dd \tau \neq 0$ only within the interval $\tau \in (0,\Delta \tau)$ and $\dd \Pi/\dd \tau \mathop{=} 0$ otherwise.
Physically, this evolution is equivalent to fully opening a half-closed valve.
The function $\Pi(\tau)$ is a polynomial of third degree in the variable $\tau/\Delta \tau$.
Fig. \ref{fig:RSSFT} shows $\Pi(\tau)$ for the U-flow defined by Table \ref{tab:FrozenTurbul} and $\Delta \tau \mathop{=}0.002$,
which reproduces quite faithfully the experimental results reported in \citep{CL+17}.

\subsection{Analytical model for Reynolds shear stress (RSS) and Reynolds-shear-stress radial gradient (RSSRG)}
\label{sct:AMRSS}

\begin{figure}[h]
	\begin{center}
		\leavevmode
		\includegraphics[width=0.6 \textwidth, trim = 0mm 0mm 0mm 0mm, clip=true]{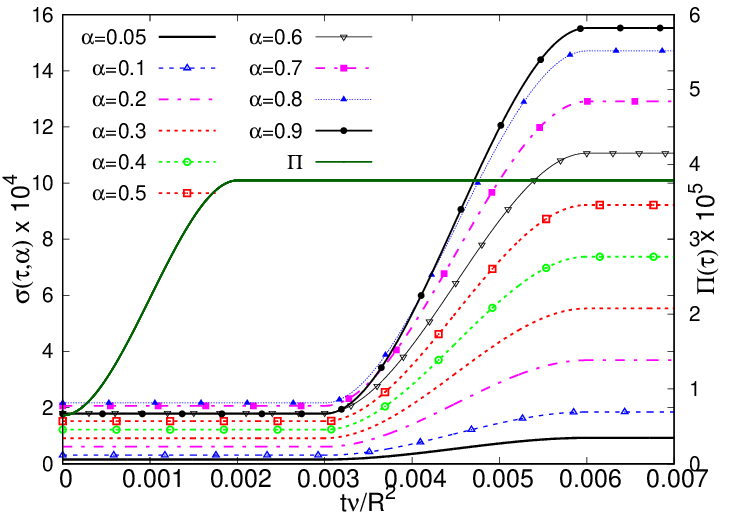}
		\caption{\small{Evolution of MPG $\Pi(\tau)$, Eq. \eqref{eq:AK+13PG}, and U-RSS $\sigma(\tau,\alpha)$, Eq. \eqref{eq:RSSG}, for the FTAM of Table \ref{tab:FrozenTurbul}, for ten radial positions ($\Delta \tau=0.002$, $\tau_0=0.003$, $\tau_2=0.006$).}}
		\label{fig:RSSFT}
	\end{center}
\end{figure}

The RSS and RSSRG functions are modelled with the Pai's polynomials that fit into Eqs. \eqref{eq:PipeSflowU} and \eqref{eq:PaiFlow}, see \citep[Sect. 7.1]{GF22},
valid for both, the initial and final S-flows:
 \begin{equation}
\label{eq:PaiRSS}
\sigma_i(\alpha)\mathop{=}\frac{\Pi_i}{2} \frac{q_i(\chi_i-1)}{\chi_i(q_i-1)}\alpha \left(1-\alpha^{2q_i-2} \right)\mathop{=}\frac{\widehat{\Pi}_i}{2} \ \alpha \left(1-\alpha^{2q_i-2} \right)
\end{equation}
 \begin{equation} \label{eq:RSSGPol}
 \varSigma_i(\alpha)\equiv  \frac{-1}{\alpha}\frac{\dd (\alpha \sigma_i)}{\dd \alpha}\mathop{=}
 - \Pi_i \frac{q_i(\chi_i-1)}{ \chi_i(q_i-1)}  \left(1- q_i \, \alpha^{2q_i-2} \right)\mathop{=}
 -\widehat{\Pi}_i \left(1- q_i\ \alpha^{2q_i-2} \right)
\end{equation}
with $q_i \in \mathbb{N}$, $q_i\mathop{\geq}2$ a best-fitting integer and $\chi_i \in \mathbb{R}$, $1\mathop{\leq} \chi_i \mathop{<}q_i$,
the CTD defined above.
Here $\widehat{\Pi}_i$ plays in the PTC an alike role to $\Pi_i$ in the ULF, and is called the \textbf{Weighted Mean-Pressure Gradient} (WMPG), being $\widehat{\Pi}_i \lesssim \Pi_i$.
The accuracy of those polynomials respect to actual S-RSS and S-RSSRG is identical to the accuracy of Eq. \eqref{eq:PaiFlow} respect to the actual S-velocity,
which was shown to be high for moderate ${Re}$.

The transient RSS and RSSRG will be modelled with a nonlinear homotopy similar to that introduced in \citep[Eq. (50)]{GF21}, which is repeated in Eq. \eqref{eq:RSSG}.
The transient RSS and RSSRG have a delay $\tau_0$ respect to the applied MPG; that is, 
$\sigma(\tau,\alpha)\mathop{=}\sigma_1(\alpha)$ for $\tau \leq \tau_0$.
This is a way of simulating that RSS and RSSRG become frozen upon changing the MPG, and start to evolve some time after the MPG began its increase, {i.e.},
after the fluid has acquired some new motion.
Thus, the TDoF $\tau_0$ would be related to the mean frozen-turbulence time (unchanged RSSRG), 
just as $\Delta \tau$ is associated to the mean valve-aperture time.
Moreover, it will be assumed that the evolution of RSS and RSSRG stops at time $\tau_2$,
which should be considered the mean turbulence-settling time,
and is the second TDoF that characterises the RSS and RSSRG functions.
Recall that all fields considered in this investigation are ensemble-averaged quantities that result after an infinite number of realisations, see \citep[Sect. 3.2.1.1]{Gar17}.
For $\tau > \tau_2$ the mean velocity would still evolve, albeit usually not much, but the RSSRG has become stationary.
The period $\widehat{\Delta \tau}\equiv \tau_2-\tau_0$ will be called the mean turbulence-settling interval, 
and plays for the RSSRG a similar role as $\Delta \tau$ for the MPG.
Thus, the proposed time-dependent RSS and RSSRG fields with two TDoF are:
\begin{align}
	\label{eq:RSSG}
	\stirlingii{\sigma(\tau,\alpha)}{\varSigma(\tau,\alpha)}\mathop{=}
	\begin{cases}
		 \stirlingii{\sigma_1(\alpha)}{\varSigma_1(\alpha)} \ \ &if \ \ \tau \leq \tau_0 \\
		(1+2\gamma^2 \left(\gamma-\tfrac{3}{2})\right) \stirlingii{\sigma_1(\alpha)}{\varSigma_1(\alpha)}-  2\ \gamma^2 \left(\gamma - \tfrac{3}{2} \right) \stirlingii{\sigma_2(\alpha)}{\varSigma_2(\alpha)}  \ \ &if \ \ \tau_0 \leq \tau \leq \tau_2 \\
		\stirlingii{\sigma_2(\alpha)}{\varSigma_2(\alpha)} \ \ &if \ \ \tau \geq \tau_2 \\
	\end{cases}
\end{align}
where 
\begin{equation}
\label{eq:gammaHom}
\gamma\mathop{=}\frac{\tau-\tau_0}{\tau_2-\tau_0}\mathop{=}\frac{\tau-\tau_0}{\widehat{\Delta \tau} } \in [0,1]
\end{equation}
is the homotopy parameter and $\sigma_i(\alpha)$, $\varSigma_i(\alpha)$ are given respectively by Eqs. \eqref{eq:PaiRSS}, \eqref{eq:RSSGPol}, 
and represent the initial and final steady-state functions.
Note $\partial \varSigma(\tau,\alpha)/\partial \tau \neq 0$ only within the interval $\tau \in (\tau_0, \tau_2)$, and $\partial \varSigma(\tau,\alpha)/\partial \tau = 0$ otherwise.
Note also that $\Delta \tau$ does not appear in Eq. \eqref{eq:RSSG}; instead $\widehat{\Delta \tau}$ does.

Fig. \ref{fig:RSSFT} shows the evolution of RSS $\sigma(\tau,\alpha)$, Eq. \eqref{eq:RSSG}, in ten radial positions, for the initial and final S-flows of Table \ref{tab:FrozenTurbul}.
The TDoF are: $\Delta \tau\mathop{=}0.002$, $\tau_0\mathop{=}0.003$, $\tau_2\mathop{=}0.006$.
Fig. \ref{fig:RSSFT} should be compared with 
\citep[Figs. 8g-h]{HSH16} (after 8 realisations), and with caution with \citep[Fig. 5]{GLC21} (after 3 realisations).
The general trend is qualitatively similar, though there are some differences.
However, it is stressed that Eq. \eqref{eq:RSSG} models a mathematical ensemble-average, made up of infinite realisations,
whereas the references' graphs are the outcome after 8 and 3 realisations, respectively.
Averaging is an operation that smoothens considerably the functions involved: the result is a much smoother function than any of its initial components.
In fact, nobody has ever seen a mathematically rigorous RSS and it can only be guessed how it would look like.
For the purpose of the present research, Eq. \eqref{eq:RSSG} is assumed sufficiently accurate and the yielded results seem to confirm such an assumption
(in particular, the outcome of TEg and TEh in Sect. \ref{sct:TEg} illustrate a noticeable coincidence with DNS friction coefficients).

Three stages are clearly distinguished in Fig. \ref{fig:RSSFT}.
Using the nomenclature reported in \citep{HS13}, namely, pre-transition, transition and fully-turbulent flow,
it may roughly be associated pre-transition with the interval $\tau \in (0,0.003)$, transition with $\tau \in (0.003,0.006)$, 
and fully-turbulent flow with $\tau \in (0.006,\infty)$.
Those three stages were already discussed in \citep[Sup-Sect. 6.3]{GF24}.

\section{Mean-velocity fields resulting from the FTAM}
\label{sct:meanVelFieldS}

With the proposed FTAM, only the TDoF $\Delta \tau$, $\tau_0$ and $\tau_2$ need be defined.
 $\Delta \tau$ is exclusively related with the laminar part of the flow (the ULF), Eq. \eqref{eq:AK+13PG},
whereas $\tau_0$ and $\tau_2$ determine the source of the PTC, Eq. \eqref{eq:RSSG}.
In the realm of the TULF, they cannot intermix. 
The mean-velocity field arising from Eq. \eqref{eq:equationGarcia} and Eqs. \eqref{eq:PaiFlow}, \eqref{eq:AK+13PG} and \eqref{eq:RSSG},
will be presented in this section.
Mathematical derivations, not previously published, can be found in the appendices.

The \textbf{IniTrans} for this U-flow is identical to that calculated in \citep[Eq. 3.12]{GF19c}, and takes the form: 
%
\begin{equation} 
 \label{eq:iniTrans}
 u_I(\tau,\alpha) =  \frac{\Pi_1}{2 \chi_1 (q_1-1)} \sum \limits_{n=1}^{\infty} \frac{J_0 (\lambda_n \alpha)}{J_1(\lambda_n)} \left[ \frac{\chi_1-1}{\lambda_n}-
 \frac{4(\chi_1-q_1)}{\lambda_n^3} - 
\frac{\chi_1-1}{2(q_1+1)J_1(\lambda_n)} {}_1F_2 \left(q_1+1;q_1+2,1;-\frac{\lambda_n^2}{4} \right)  \right] \euler^{-\lambda_n^2 \tau}
\end{equation}
with ${}_1F_2(q_i+1;q_i+2,1;-\lambda_n^2/4)\equiv {}_1F_2[i], \ i=1,2,$ the generalised hypergeometric function, see \citep[Ch. 15]{AS72}.
It can be further decomposed into ULF and PTC components of IniTrans, yielding an ULF:
\begin{equation}
 \label{eq:iniTransULF}
 u_{I_L}(\tau,\alpha) = 2\Pi_1 \sum \limits_{n=1}^{\infty} \frac{J_0 (\lambda_n \alpha)}{\lambda_n^3 J_1(\lambda_n)} \ \euler^{-\lambda_n^2 \tau}
\end{equation}
and a PTC:
\begin{equation}
 \label{eq:iniTransPTC}
 u_{I_T}(\tau,\alpha) = \frac{\widehat{\Pi}_1}{2 q_1} \sum \limits_{n=1}^{\infty} \frac{J_0 (\lambda_n \alpha)}{J_1(\lambda_n)} \left[ \frac{1}{\lambda_n}-
 \frac{4q_1}{\lambda_n^3} -  \frac{ {}_1F_2[1]}{2(q_1+1)J_1(\lambda_n)}   \right] \euler^{-\lambda_n^2 \tau}
\end{equation}
Note the PTC becomes zero if $\chi_1=1$, which eventually is the TULF's standard procedure to remove the turbulence contribution.

The \textbf{PresGrad} component stems from the source $\Pi(\tau)$ of Eq. \eqref{eq:AK+13PG} and is somewhat more involved.
Its derivation can be found in App. \ref{app:DerPresGrad}.
PresGrad takes two expressions, according to $\tau < \Delta \tau$ or  $\tau > \Delta \tau$:
%
\begin{align}
 \label{eq:PresGradBefore}
 & u_P(\tau,\alpha) = \frac{2(\Pi_1-\Pi_2)}{\Delta \tau^2} \sum \limits_{n=1}^{\infty} \frac{J_0 (\lambda_n \alpha)}{\lambda_n^3 J_1(\lambda_n)} 
 \left[ \frac{2 \tau^3}{\Delta \tau}-
3 \left( \frac{2}{\Delta \tau \lambda_n^2} +1 \right) \tau^2 + \left(\frac{2}{\Delta \tau \lambda_n^2 } +1 \right)\frac{6\tau}{\lambda_n^2} - \right.  \nonumber \\
 & \left. 6 \left( \frac{2}{\Delta \tau \lambda_n^6} + \frac{1}{\lambda_n^4} - \frac{\Pi_1 \ \Delta \tau^2}{6 (\Pi_1-\Pi_2)} \right) \left( 1- \euler^{-\lambda_n^2 \tau} \right)  \right], \qquad (0 \leq \tau \leq \Delta \tau)
\end{align}
\begin{align}
 \label{eq:PresGradAfter}
 & u_P(\tau,\alpha) = \frac{2(\Pi_1-\Pi_2)}{\Delta \tau^2} \sum \limits_{n=1}^{\infty} \frac{J_0 (\lambda_n \alpha)}{\lambda_n^3 J_1(\lambda_n)} 
 \left[ \frac{\Pi_2 \ \Delta \tau^2}{\Pi_1-\Pi_2} +
\left(1- \frac{2}{\Delta \tau \lambda_n^2} \right)\frac{6 \euler^{-\lambda_n^2(\tau-\Delta \tau)}}{\lambda_n^4} + \right.  \nonumber \\
 & \left. 6 \left( \frac{2}{\Delta \tau \lambda_n^6} + \frac{1}{\lambda_n^4} - \frac{\Pi_1 \ \Delta \tau^2}{6 (\Pi_1-\Pi_2)} \right) \euler^{-\lambda_n^2 \tau}  \right], \qquad ( \tau \geq \Delta \tau)
\end{align}
Note that only the DoF $\Pi_i$ and $\Delta \tau$ appear in these equations.
Other DoF, such as $q_i$, $\chi_i$, $\tau_0$ or $\tau_2$, do not appear, because they are related with turbulence.

The \textbf{RStress} component arising from source $\varSigma(\tau,\alpha)$ of Eq. \eqref{eq:RSSG} is the most complicated of all.
Its derivation can be found in App. \ref{app:DerRStress}.
RStress takes three expressions, according to $\tau < \tau_0$, $\tau_0 \leq \tau \leq \tau_2$ or  $\tau > \tau_2$:
\begin{align}
 \label{eq:RStressBefore}
 u_R(\tau,\alpha) = \widehat{\Pi}_1  \sum \limits_{n=1}^{\infty} \frac{J_0 (\lambda_n \alpha)}{ [\lambda_n J_1(\lambda_n)]^2} & \left[ {}_1F_2[1] - \frac{2 J_1(\lambda_n)}{\lambda_n} \right] \left[ 1-\euler^{-\lambda_n^2 \tau} \right] \qquad (0 \leq \tau \leq \tau_0)
\end{align}
\begin{align}
 \label{eq:RStressMiddle}
 & u_R(\tau,\alpha) = \sum \limits_{n=1}^{\infty} \frac{J_0 (\lambda_n \alpha)}{[\lambda_n J_1(\lambda_n)]^2} \left \{ \widehat{\Pi}_1 \left[ {}_1F_2[1] - \frac{2 J_1(\lambda_n)}{\lambda_n} \right] \left(1-\euler^{-\lambda_n^2 \tau} \right) 
  +\frac{6}{\widehat{\Delta \tau}^3}\left[ \frac{2J_1(\lambda_n)}{\lambda_n} (\widehat{\Pi}_2-\widehat{\Pi}_1)+ {}_1F_2[1] \widehat{\Pi}_1- {}_1F_2[2]\widehat{\Pi}_2  \right] \right. \nonumber \\ 
& \times \left[ \frac{(\tau-\tau_0)^3}{3} -\left(\frac{2}{\lambda_n^2}+\widehat{\Delta \tau} \right) \left[ \frac{(\tau-\tau_0)^2}{2}-\frac{\tau-\tau_0}{\lambda_n^2} + 
 \left. \frac{1}{\lambda_n^4} \left(1-\euler^{-\lambda_n^2(\tau-\tau_0)}  \right)\right]\right] \right \}, \qquad (\tau_0 \leq \tau \leq \tau_2)
\end{align}
\begin{align}
 \label{eq:RStressAfter}
 & u_R(\tau,\alpha) = \sum \limits_{n=1}^{\infty} \frac{J_0 (\lambda_n \alpha)}{[\lambda_n J_1(\lambda_n)]^2} \left \{ \widehat{\Pi}_1 \left[ {}_1F_2[1] - \frac{2 J_1(\lambda_n)}{\lambda_n} \right] 
  \left(\euler^{-\lambda_n^2 (\tau-\tau_2)}-\euler^{-\lambda_n^2 \tau} \right) + \widehat{\Pi}_2 \left[ {}_1F_2[2] - \frac{2 J_1(\lambda_n)}{\lambda_n} \right]\left[1-\euler^{-\lambda_n^2 (\tau-\tau_2)}\right] \right. \nonumber \\
 &\left . +6\left[ \frac{2J_1(\lambda_n)}{\lambda_n}(\widehat{\Pi}_2-\widehat{\Pi}_1)+{}_1F_2[1] \widehat{\Pi}_1-{}_1F_2[2] \widehat{\Pi}_2 \right] 
  \left[ \left(\frac{-1}{6}+\frac{1}{\widehat{\Delta \tau}^2 \lambda_n^4 }-\frac{2}{ \widehat{\Delta \tau}^3 \lambda_n^6}\right) \euler^{-\lambda_n^2 (\tau-\tau_2)}  + \frac{2+\widehat{\Delta \tau}\lambda_n^2}{\widehat{\Delta \tau}^3 \lambda_n^6} \ \euler^{-\lambda_n^2(\tau-\tau_0)} \right] \right \} \nonumber \\
&  \qquad \qquad \qquad \qquad \qquad \qquad \qquad \qquad \qquad \qquad  \qquad \qquad \qquad \qquad \qquad \qquad  (\tau \geq \tau_2)
\end{align}
Note that only the DoF $\widehat{\Pi}_i$, $q_i$, $\tau_0$, $\tau_2$ and $\widehat{\Delta \tau}$ appear in these equations, 
because they are related with turbulence.

 \subsection{The bulk velocity and mean wall-shear stress (MWSS)}
 
 \label{sct:BulkWSS}

 The bulk velocity for mean pipe flow is given by 
 \begin{equation}
  \label{eq:bulkSeries}
  \widetilde{u}(\tau) \mathop{=} 2 \int \limits_0^1 \alpha \ u(\tau,\alpha) \dd \alpha= 2 \sqrt{2} \sum \limits_{n=1}^{\infty} \frac{ u_{I_n}(\tau)+u_{P_n}(\tau)+u_{R_n}(\tau)} {\lambda_n}
 \end{equation}
where $u_{I_n}(\tau)$, $u_{P_n}(\tau)$ and $u_{R_n}(\tau)$ equals all that is multiplying $\phi_n(\alpha)=\sqrt{2}J_0(\lambda_n \alpha)/J_1(\lambda_n)$ in the series of Eqs. \eqref{eq:iniTrans}-\eqref{eq:RStressAfter}.

The mean wall-shear stress (MWSS) is defined as
\begin{equation}
 \label{eq:WSSsymmet}
 \sigma_w(\tau) \mathop{=}\frac{\partial u(\tau,1)}{\partial \alpha} \mathop{=} -\sqrt{2} \sum \limits_{n=1}^{\infty} \lambda_n \left[ u_{I_n}(\tau) + u_{P_n}(\tau)+ u_{R_n}(\tau) \right ]
\end{equation}
Eqs. \eqref{eq:bulkSeries} and \eqref{eq:WSSsymmet} will be profusely used to calculate the curves reported in this research, particularly those of $C_f(\tau)$.

\section{Derivation of PresGrad component {$u_P(\tau,\alpha)$}}\label{app:DerPresGrad}

The following integrals are necessary:
\begin{equation}
\label{eq:tcuadrado}
 \int \limits_0^{\tau} \tau'^2 \euler^{\lambda_n^2 \tau'} \dd \tau'= \frac{1}{\lambda_n^6} \left[  \left( \lambda_n^4 \tau^2 - 2 \lambda_n^2 \tau + 2 \right) \euler^{\lambda_n^2 \tau} -2 \right]
\end{equation}
\begin{equation}
\label{eq:tcubo}
 \int \limits_0^{\tau} \tau'^3 \euler^{\lambda_n^2 \tau'} \dd \tau'= \frac{1}{\lambda_n^8} \left[  \left( \lambda_n^6 \tau^3 -3 \lambda_n^4 \tau^2+ 6 \lambda_n^2 \tau -6 \right) \euler^{\lambda_n^2 \tau} +6 \right]
\end{equation}
For $\tau < \Delta \tau$, the integral of $\Pi(\tau)$, Eq. \eqref{eq:AK+13PG}, as demanded in Eq. \eqref{eq:equationGarcia} becomes:
%
\begin{align}
& \int \limits_0^{\tau} \Pi(\tau') \euler^{-\lambda_n^2 (\tau-\tau')} \dd \tau'= \euler^{-\lambda_n^2 \tau}  \int \limits_0^{\tau}\left( \frac{2(\Pi_1-\Pi_2)}{\Delta \tau^3} \tau'^3 - 
 \frac{3(\Pi_1-\Pi_2)}{\Delta \tau^2} \tau'^2 + \Pi_1   \right)\euler^{\lambda_n^2 \tau'} \dd \tau' = \nonumber \\
 & \frac{2(\Pi_1-\Pi_2)}{\Delta \tau^3 \lambda_n^2}\tau^3- \frac{3(\Pi_1-\Pi_2)}{\Delta \tau^2 \lambda_n^2} \left( \frac{2}{\Delta \tau \lambda_n^2} +1 \right) \tau^2+ 
  \frac{6(\Pi_1-\Pi_2)}{\Delta \tau^2 \lambda_n^4} \left( \frac{2}{\Delta \tau \lambda_n^2} +1 \right) \tau - \nonumber \\
  &\left( \frac{12(\Pi_1-\Pi_2)}{\Delta \tau^3 \lambda_n^6} + \frac{6(\Pi_1-\Pi_2)}{\Delta \tau^2 \lambda_n^4} -\Pi_1\right) \frac{1-\euler^{-\lambda_n^2 \tau}}{\lambda_n^2}\ , \qquad \qquad (0\leq \tau <\Delta \tau)
\end{align}
which after some algebra yields Eq. \eqref{eq:PresGradBefore}.
For $\tau \geq \Delta \tau$ the result is:
%
\begin{align}
& \int \limits_0^{\tau} \Pi(\tau') \euler^{-\lambda_n^2 (\tau-\tau')} \dd \tau' \mathop{=}
\euler^{-\lambda_n^2 \tau} \int \limits_0^{\Delta \tau} \Pi(\tau') \euler^{\lambda_n^2 \tau'} \dd \tau' + 
\Pi_2 \ \euler^{-\lambda_n^2 \tau} \int \limits_{\Delta \tau}^{\tau} \euler^{\lambda_n^2 \tau'} \dd \tau'=  \nonumber \\
& \frac{\euler^{-\lambda_n^2 (\tau-\Delta \tau)}}{\lambda_n^2 }\left[- \frac{12(\Pi_1-\Pi_2)}{\Delta \tau^3 \lambda_n^6} + \frac{6(\Pi_1-\Pi_2)}{\Delta \tau^2 \lambda_n^4} +\Pi_2\right]  + \nonumber \\
& \frac{\euler^{-\lambda_n^2 \tau}}{\lambda_n^2 }\left[ \frac{12(\Pi_1-\Pi_2)}{\Delta \tau^3 \lambda_n^6} + \frac{6(\Pi_1-\Pi_2)}{\Delta \tau^2 \lambda_n^4} -\Pi_1 \right] \mathop{+} \Pi_2 \frac{1- \euler^{-\lambda_n^2 (\tau-\Delta \tau)}}{\lambda_n^2}  \ , \qquad \qquad (\tau \geq \Delta \tau)
\end{align}
which after some algebra yields Eq. \eqref{eq:PresGradAfter}.\\

  
\section{Derivation of RStress component {$u_R(\tau,\alpha)$}} \label{app:DerRStress}

In the Hilbert space $L^2_{\alpha}[0,1]$ of solutions of the governing equation, it is defined an inner product $\varSigma_n\mathop{=}\langle \varSigma,\phi_n \rangle$, given by   
\begin{equation}
 \label{eq:baseH}
  \langle \varSigma ,\phi_n \rangle = \int \limits_0^1 \alpha\ \varSigma(\tau,\alpha) \phi_n(\alpha) \dd\alpha \ , \qquad \phi_n(\alpha) = \frac{\sqrt{2}J_0(\lambda_n \alpha)}{J_1(\lambda_n)}
\end{equation}
From Eqs. \eqref{eq:RSSGPol}-\eqref{eq:RSSG}, it follows that the expression of such an inner product, 
which is needed to calculate the terms containing $\varSigma_n$ in Eq. \eqref{eq:equationGarcia}, will include the following integrals:

\begin{equation}\label{eq:intJ0}
 \int \limits_0^1 \alpha J_0(\lambda_n\alpha) \dd \alpha = \frac{J_1(\lambda_n)}{\lambda_n}
\end{equation}
\begin{equation}\label{{eq:intAlphaqJ0}}
  \int \limits_0^1 \alpha^{2q_i-1} J_0(\lambda_n\alpha) \dd \alpha = \frac{1}{2q_i}{}_1F_2(q_i;q_i+1,1;-\lambda_n^2/4) \equiv \frac{1}{2q_i}{}_1F_2[i]\ , \qquad (i=1,2)   
\end{equation}
with ${}_1F_2(a;b,c;z)$ the generalised hypergeometric function, see \citep[Ch. 15]{AS72}.
Accordingly, the inner product for different time intervals results as:
%
\begin{equation}
\label{eq:varsigmaLow}
 \varSigma_n\mathop{=}\langle \varSigma_1,\phi_n \rangle\mathop{=}\mathop{-}\int \limits_0^1 \alpha \widehat{\Pi}_1 \left(1 \mathop{-} q_1 \alpha^{2q_1-2} \right)\frac{\sqrt{2}J_0(\lambda_n \alpha)}{J_1(\lambda_n)} \dd \alpha\mathop{=}\frac{ \widehat{\Pi}_1}{\sqrt{2}\ J_1(\lambda_n)}\left( {}_1F_2[1]\mathop{-}\frac{2\ J_1(\lambda_n)}{\lambda_n} \right) , \  (0\leq \tau < \tau_0)
\end{equation}
\begin{align}
\label{eq:varsigmaMiddle}
&\varSigma_n(\tau)\mathop{=}-\int \limits_0^1 \left[ \widehat{\Pi}_1\left( \alpha-q_1 \alpha^{2q_1-1} \right) \left(1\mathop{+}2\gamma^2(\gamma-\tfrac 32) \right) -\widehat{\Pi}_2 \left(\alpha-q_2 \alpha^{2q_2-1} \right)2\gamma^2(\gamma - \tfrac 32) \right] \frac{\sqrt{2}\ J_0(\lambda_n \alpha)}{J_1(\lambda_n)} \dd \alpha \mathop{=} \nonumber \\
&\frac{\widehat{\Pi}_1}{\sqrt{2}\ J_1(\lambda_n)}\left( {}_1F_2[1]-\frac{2\ J_1(\lambda_n)}{\lambda_n} \right)\mathop{+}\frac{2\gamma^2(\gamma-\tfrac 32)}{\sqrt{2}\ J_1(\lambda_n)}\left( \frac{2\ J_1(\lambda_n)}{\lambda_n}(\widehat{\Pi}_2-\widehat{\Pi}_1)\mathop{+} {}_1F_2[1]\ \widehat{\Pi}_1- {}_1F_2[2]\  \widehat{\Pi}_2 \right)  , \ (\tau_0 \leq \tau \leq \tau_2)
\end{align}
\begin{equation}
\label{eq:varsigmaHigh}
 \varSigma_n\mathop{=}\langle \varSigma_2,\phi_n \rangle\mathop{=}\frac{ \widehat{\Pi}_2}{\sqrt{2}\ J_1(\lambda_n)}\left( {}_1F_2[2]-\frac{2\ J_1(\lambda_n)}{\lambda_n} \right) \ , \qquad (\tau > \tau_2)
\end{equation}
with $\gamma\mathop{\mathop{=}}(\tau- \tau_0)/(\tau_2-\tau_0)\mathop{=}(\tau- \tau_0)/\widehat{\Delta \tau}$ defined by Eq. \eqref{eq:gammaHom}.
The time integral of $\varSigma_n$ in Eq. \eqref{eq:equationGarcia} is calculated using Eqs. \eqref{eq:varsigmaLow}-\eqref{eq:varsigmaHigh} and \eqref{eq:tcuadrado}-\eqref{eq:tcubo}:
\begin{align}
\int \limits_0^{\tau} \varSigma_n(\tau') \euler^{-\lambda_n^2 (\tau-\tau')} \dd \tau' \mathop{=}
\euler^{-\lambda_n^2 \tau} \int \limits_0^{\tau} \frac{\widehat{\Pi}_1}{\sqrt{2} \ J_1(\lambda_n)} \left( {}_1F_2[1]-\frac{2\ J_1(\lambda_n)}{\lambda_n} \right) \euler^{\lambda_n^2 \tau'} \dd \tau' = &
 \frac{\widehat{\Pi}_1}{\sqrt{2} \ \lambda_n^2 J_1(\lambda_n)} \left( {}_1F_2[1]-\frac{2\ J_1(\lambda_n)}{\lambda_n} \right) \left( 1-\euler^{-\lambda_n^2 \tau} \right)\ , \nonumber \\
& \qquad \qquad  \qquad \qquad \qquad \qquad  (0\leq \tau < \tau_0)
\end{align}
\begin{align}
&\int \limits_0^{\tau} \varSigma_n(\tau') \euler^{-\lambda_n^2 (\tau-\tau')} \dd \tau' \mathop{=}
\euler^{-\lambda_n^2 \tau} \left (  \int \limits_0^{\tau_0} \varSigma_n(\tau') \euler^{\lambda_n^2 \tau'} \dd \tau' \mathop{+}  \int \limits_{\tau_0}^{\tau} \varSigma_n(\tau') \euler^{\lambda_n^2 \tau'} \dd \tau'    \right) = \nonumber \\
&\frac{\widehat{\Pi}_1}{\sqrt{2} \lambda_n^2 J_1(\lambda_n)} \left( {}_1F_2[1]-\frac{2\ J_1(\lambda_n)}{\lambda_n} \right) \left(\euler^{-\lambda_n^2 (\tau-\tau_0)}- \euler^{-\lambda_n^2 \tau} \right) \mathop{+} \euler^{-\lambda_n^2 \tau}  \int \limits_{\tau_0}^{\tau} \varSigma_n(\tau') \euler^{\lambda_n^2 \tau'} \dd \tau'   = \nonumber \\
& \frac{\widehat{\Pi}_1}{\sqrt{2} \ \lambda_n^2 J_1(\lambda_n)} \left( {}_1F_2[1]-\frac{2\ J_1(\lambda_n)}{\lambda_n} \right) \left( 1-\euler^{-\lambda_n^2 \tau} \right) \mathop{+}\frac{3 \sqrt{2}}{\lambda_n^2 J_1(\lambda_n) \widehat{\Delta \tau}^3} \left(\frac{2 J_1(\lambda_n) (\widehat{\Pi}_2-\widehat{\Pi}_1)}{\lambda_n}\mathop{+} {}_1F_2[1] \widehat{\Pi}_1 - {}_1F_2[2]\widehat{\Pi}_2 \right)  \nonumber \\
&\times \left[ \frac{(\tau-\tau_0)^3}{3}-\frac{2\mathop{+}\lambda_n^2\widehat{\Delta \tau}}{2\lambda_n^2}(\tau-\tau_0)^2\mathop{+}\frac{2+\lambda_n^2\widehat{\Delta \tau}}{\lambda_n^4}(\tau-\tau_0)-\frac{2+\lambda_n^2\widehat{\Delta \tau}}{\lambda_n^6}+\frac{2+\lambda_n^2\widehat{\Delta \tau}}{\lambda_n^6}\euler^{-\lambda_n^2(\tau-\tau_0)}  \right], \qquad (\tau_0\leq \tau \leq \tau_2)
\end{align}
\begin{align}
&\int \limits_0^{\tau} \varSigma_n(\tau') \euler^{-\lambda_n^2 (\tau-\tau')} \dd \tau' \mathop{=}
\euler^{-\lambda_n^2 \tau} \left (  \int \limits_0^{\tau_0} \varSigma_n(\tau') \euler^{\lambda_n^2 \tau'} \dd \tau' +  \int \limits_{\tau_0}^{\tau_2} \varSigma_n(\tau') \euler^{\lambda_n^2 \tau'} \dd \tau' +  \int \limits_{\tau_2}^{\tau} \varSigma_n(\tau') \euler^{\lambda_n^2 \tau'} \dd \tau'   \right) = \nonumber \\
&\euler^{-\lambda_n^2 \tau} \left \{ \frac{\widehat{\Pi}_1}{\sqrt{2} \lambda_n^2 J_1(\lambda_n)} \left( {}_1F_2[1]-\frac{2\ J_1(\lambda_n)}{\lambda_n} \right) \left(\euler^{\lambda_n^2\tau_2}- 1 \right) +
\frac{3\sqrt{2}\ \euler^{\lambda_n^2 \tau_0}}{\lambda_n^2 J_1(\lambda_n) \widehat{\Delta \tau}^3} \left(\frac{2 J_1(\lambda_n) (\widehat{\Pi}_2-\widehat{\Pi}_1)}{\lambda_n}+ {}_1F_2[1] \widehat{\Pi}_1 - {}_1F_2[2]\widehat{\Pi}_2 \right) \right. \nonumber \\
&\times \left[ \left( \frac{\widehat{\Delta \tau}^3}{3}-\frac{\widehat{\Delta \tau}^2}{\lambda_n^2}+\frac{2\widehat{\Delta \tau}}{\lambda_n^4}-\frac{2}{\lambda_n^6}\right)\euler^{\lambda_n^2 \widehat{\Delta \tau}}+ \frac{2}{\lambda_n^6} - \left( \frac{\widehat{\Delta \tau}^3}{2}-\frac{\widehat{\Delta \tau}^2}{\lambda_n^2} + \frac{\widehat{\Delta \tau}}{\lambda_n^4} \right)\euler^{\lambda_n^2 \widehat{\Delta \tau}}+\frac{\widehat{\Delta \tau}}{\lambda_n^4} \right]+ \nonumber \\
& \left. \frac{\widehat{\Pi}_2}{\sqrt{2} \lambda_n^2 J_1(\lambda_n)} \left( {}_1F_2[2]-\frac{2\ J_1(\lambda_n)}{\lambda_n} \right) \left(\euler^{\lambda_n^2\tau}- \euler^{\lambda_n^2\tau_2} \right) \right \}  = \frac{1}{\sqrt{2} \lambda_n^2 J_1(\lambda_n)}\left \{ \widehat{\Pi}_1 \left( {}_1F_2[1]-\frac{2\ J_1(\lambda_n)}{\lambda_n} \right) \left(\euler^{-\lambda_n^2(\tau-\tau_2)}- \euler^{-\lambda_n^2\tau} \right)\right.  \nonumber \\
& +\widehat{\Pi}_2 \left( {}_1F_2[2]-\frac{2\ J_1(\lambda_n)}{\lambda_n} \right) \left(1-\euler^{-\lambda_n^2(\tau-\tau_2)} \right)+ \frac{6}{\widehat{\Delta \tau}^3}\left(\frac{2 J_1(\lambda_n) (\widehat{\Pi}_2-\widehat{\Pi}_1)}{\lambda_n}+ {}_1F_2[1] \widehat{\Pi}_1 - {}_1F_2[2]\widehat{\Pi}_2 \right) \euler^{-\lambda_n^2(\tau-\tau_0)} \times \nonumber \\
&\left. \left[ \left(-\frac{\widehat{\Delta \tau}^3}{6} + \frac{\widehat{\Delta \tau}}{\lambda_n^4 } - \frac{2}{\lambda_n^6} \right)\euler^{\lambda_n^2 \widehat{\Delta \tau}}+\frac{2+\lambda_n^2 \widehat{\Delta \tau}}{\lambda_n^6} \right] \right \}, \qquad \qquad (\tau \geq \tau_2)
\end{align}
With some algebra, these integrals yield directly Eqs. \eqref{eq:RStressBefore}-\eqref{eq:RStressAfter}, respectively.

\end{document}